\documentclass[12pt]{article}
\usepackage{amsmath,amssymb}
\usepackage{epsfig}

\begin{document}
\numberwithin{equation}{section}
\newtheorem{theorem}{Theorem}[section]
\newtheorem{proposition}{Proposition}[section]
\newtheorem{remark}{Remark}[section]
\newtheorem{lemma}{Lemma}[section]
\newtheorem{corollary}{Corollary}
\newtheorem{definition}{Definition}[section]
\pagestyle{headings}
\def\NN{{\rm I\hspace{-0.50ex}N} }
\def\ZZ{{\sf Z\hspace{-0.80ex}Z} }
\def\RR{{\rm I\hspace{-0.50ex}R} }
\def\PP{{\rm I\hspace{-0.50ex}P} }
\def\EE{{\rm I\hspace{-0.50ex}E} }
\def\CC{\rm \hbox{C\kern-.57em\raise.47ex
         \hbox{$\scriptscriptstyle |$}\kern+0.5 em }}
\def\II{{1\hspace{-0.70ex}\rm I} }

\newcommand \dps {\displaystyle }

\newcommand \tr {{\rm Tr}}

\newcommand{\mc}{{\mathcal C}}
\newcommand{\me}{{\mathcal E}}
\newcommand{\mm}{{\mathcal M}}
\newcommand{\mo}{{\mathcal O}}
\newcommand{\muu}{{\mathcal U}}

\title{Multilevel domain decomposition for electronic structure calculations}
\author{M.~Barrault$^1$, E. ~Canc\`es$^2$,  W.~W. Hager $^3$ and C. Le Bris$^2$\\
{\footnotesize $^1$ EDF R\&D, 1 avenue du G\'en\'eral de Gaulle, 92141
  Clamart Cedex, France}\\ 
{\footnotesize
  \tt{maxime.barrault@edf.fr}}\\
{\footnotesize $^2$ CERMICS, \'Ecole Nationale des Ponts et
Chauss\'ees,}\\
{\footnotesize 6 \& 8, avenue Blaise Pascal, Cit\'e Descartes,}\\
{\footnotesize  77455 Marne-La-Vall\'ee Cedex 2, France \tt{\{cances,lebris\}@cermics.enpc.fr}}\\
{\footnotesize $^3$ Department of Mathematics, University of Florida,}\\
{\footnotesize
  Gainesville FL 32611-8105, USA, \tt{hager@math.ufl.edu}}
}
\maketitle

\begin{abstract}
We introduce a new multilevel domain decomposition method (MDD) for
electronic structure 
calculations within semi-empirical and Density Functional Theory (DFT)
frameworks. This method iterates between local fine solvers and 
global coarse solvers, in the spirit of domain decomposition 
methods. Using this approach, calculations have been
successfully performed on several linear polymer chains
containing up to 40,000 atoms and 200,000 atomic orbitals. Both the
computational cost and the memory requirement scale linearly with the
number of atoms. Additional speed-up can easily be obtained
by parallelization. We show that this domain decomposition
method outperforms the Density Matrix Minimization (DMM) method for poor
initial guesses. Our method provides an efficient preconditioner for DMM
and other linear scaling methods, variational in nature, such as the
Orbital Minimization (OM) procedure.  
\end{abstract}

\newpage

%%%%%%%%%%%%%%%%%%%%%%%%%%%%%%%%%%%%%%%%%%%%%%%%%%%%%%%%%%%%%%%%%%%%%%
\section{Introduction and motivation}

A central issue in computational quantum chemistry is the
determination of the electronic ground state of a
molecular system. For completeness and self-consistency, we now briefly
introduce the problem. In particular, we present it in a mathematical
way.  

\subsection{Standard electronic structure calculations}

A molecular system is composed of $N$ electrons,
modelled quantum mechanically,  and a given number of nuclei,
the latter being considered as classical point-like particles clamped at
known positions (Born-Oppenheimer approximation). We refer to
\cite{cances-handbook} for a general mathematical exposition and to
\cite{Hehre,McWeeny1} for the chemical background. 
Determining the electronic ground state amounts to solving a
time-independent Schr\"odinger equation in $\RR^{3N}$. 
This goal is out of reach for large values of $N$. In fact it is already
infeasible for
values of $N$ exceeding three or four, unless dedicated techniques are
employed. Examples are stochastic-like techniques such as Diffusion
Monte-Carlo approaches, or emerging 
techniques, such as sparse tensor products
techniques~\cite{acta}. Approximations of the Schr\"odinger equation have
been developed, such as the widely used \emph{tight-binding},
\emph{Hartree-Fock} 
and \emph{Kohn-Sham} models. For these three models, the
numerical resolution of a problem of the
following type is required: given $H$ and $S$, respectivement an 
$N_b \times N_b$ symmetric  matrix and an $N_b \times N_b$ 
symmetric positive definite matrix (with $N_b > N$), compute a solution
$D_\star$ of the problem 
\begin{equation}
  \label{eq:euler}
 \left\{
\begin{array}{l}
\displaystyle H c_i = \epsilon_i S c_i, 
\qquad \qquad
\displaystyle \epsilon_1 \leq \ldots \leq\epsilon_N \le \epsilon_{N+1}\leq
\ldots \leq \epsilon_{N_b},\\
\\
\displaystyle c_i^t Sc_j = \delta_{ij},\\
\\
D_\star = \dps \sum_{i=1}^N c_ic_i^t.
\end{array} \right.
\end{equation}
Let us mention that most electronic structure calculations are performed
with closed shell models \cite{Hehre}, and that, consequently, the
integer $N$ in (\ref{eq:euler}) then is the number of electron pairs.  
We remark that when $S$ is the identity matrix, a solution $D_\star$
to~(\ref{eq:euler}) is a solution to the problem
\begin{equation}
  \label{eq:pb}
\left\{
  \begin{array}{l}
\hbox{\rm  Find the {\em orthogonal projector} on
the space spanned by the}\, N\,\hbox{\rm 
eigenvectors}\\
\hbox{\rm  associated with the lowest}\, N\,\hbox{\rm  eigenvalues of $H$.}
  \end{array}
\right.
\end{equation}
In (\ref{eq:pb}), and throughout this article, the eigenvalues are counted with
 their multiplicities. The $N$ eigenvectors $c_i$, called
 \emph{generalized} eigenvectors in order to
emphasize  the presence of the matrix $S$, represent the expansion in a
 given Galerkin basis 
 $\left\{ \chi_i \right\}_{1 \le i \le N_b}$ of the $N$ one-electron
wavefunctions. The matrix $H$ is a mean-field Hamiltonian matrix. For
instance, for the Kohn-Sham model, we have
\begin{equation} \label{eq:defH}
H_{ij} = \frac 1 2 \int_{\RR^3} \nabla \chi_i \cdot \nabla
\chi_j + \int_{\RR^3} V \chi_i \chi_j
\end{equation}
where $V$ is a mean-field local potential. The matrix $S$ is the overlap
matrix associated with the basis 
 $\left\{ \chi_i \right\}_{1 \le i \le N_b}$: 
\begin{equation} \label{eq:defS}
S_{ij} = \dps \int_{\RR^3} \chi_i \chi_j.
\end{equation}
In this article, we focus on the \emph{Linear Combination of Atomic
  Orbitals} (LCAO) approach. This is a very efficient discretization
  technique, using localized basis functions $\left\{ \chi_i
  \right\}$, compactly supported~\cite{SIESTA} or exhibiting a gaussian
  fall-off~\cite{Hehre}. 

It is important to emphasize what makes the electronic structure
problem, discretized with the LCAO approach, 
specific as compared to other linear eigenvalue problems encountered in other
fields of the engineering sciences (see~\cite{Lehoucq,Lehoucq2} for
instance). First, $N_b$ is proportional to $N$, 
and not much larger than it (say $N_b \sim 2N$ to fix the ideas). Hence,
the problem is not finding a few eigenvectors of the generalized
eigenvalue problem~(\ref{eq:euler}). Second, 
although the matrices $H$ and $S$ are sparse for large molecular systems 
(see section~\ref{sec:linearscaling} for details), they are not as
sparse as the stiffness 
and mass matrices usually encountered when using finite difference or finite
element methods. For example, the bandwith of $H$ and $S$ is of the
order of $10^2$ in the numerical examples reported in
section~\ref{sec:numerical}. Note that, in contrast, for plane wave basis set
discretizations (which will not be discussed here), the parameter $N_b$
is much larger than $N$ (say $N_b \sim\, 100\, N$), the matrix $S$ is
the identity matrix and the matrix $H$ est full.
Third, and this is a crucial point, the output of the
calculation is the matrix $D_\star$ and not the generalized eigenvectors
$c_i$ themselves. This is the fundamental remark allowing the
construction of linear scaling methods (see~section~\ref{sec:linearscaling}).

\medskip

A solution $D_\star$ of~(\ref{eq:euler}) is
\begin{equation}
\label{eq:D}
\displaystyle D_\star = C_\star C_\star^t 
\end{equation}
where $C_\star$ is a solution to
the minimization problem
\begin{equation}
  \label{eq:infF}
 \inf \biggl\{\mbox{\textrm{Tr}}\Big(HCC^{t}\Big), \quad C\in
\mm^{N_{b},N}(\RR), \; C^{t}SC=I_{N} \biggr\}.
\end{equation}
Note that the energy functional $\mbox{\textrm{Tr}}\Big(HCC^{t}\Big)$  
can be given the more symmetric form
$\mbox{\textrm{Tr}}\Big(C^{t}HC\Big)$. 
Here and below, $\mm^{k,l}$ denotes the vector space of the $k \times l$
real matrices.
Notice that (\ref{eq:infF}) has many
minimizers: if $C_\star$ is a minimizer, so is $C_\star U$ for any
orthogonal $N \times N$ matrix $U$. However, under the standard
assumption that the $N$-th 
eigenvalue of $H$ is strictly lower than the $(N+1)$-th one, the
matrix $D^\ast$ defined by (\ref{eq:D}) does not in 
fact depend on the choice of the  minimizer $C_\star$ of
(\ref{eq:infF}). Notice also that (\ref{eq:euler}) are not the
Euler-Lagrange equations of (\ref{eq:infF}) but that any critical point
of (\ref{eq:infF}) is obtained from a solution of (\ref{eq:euler}) by an
orthogonal transformation of the columns of $C_\star  = \dps
\left( c_1 | \cdots |c_N \right)$.

\medskip

The standard approach to compute $D_\star$ is
to solve the generalized eigenvalue problem~(\ref{eq:euler})
and then construct $C_\star$ thus $D_\star$ by 
collecting the lowest $N$ generalized eigenvectors of $H$. This approach
is employed when the
number $N$ of electrons (or electron pairs) is not too large, say
smaller than $10^3$. 

\medskip

\subsection{Linear scaling methods}
\label{sec:linearscaling}

One of the current challenges of Computational Chemistry is to lower the
computational complexity $N^3$ of this solution procedure. A
linear complexity $N$ is the holy grail. There are various existing
methods designed for this purpose. 
Surveys on such methods are \cite{revON,FOE}. Our purpose here
is to introduce a new method, based on the \emph{domain decomposition}
paradigm. We remark that the method introduced here
is not the first occurrence of a method based on a decomposition of the
matrix $H$ \cite{DCchem}, but a significant methodological
  improvement is fulfilled with the present method.
 To the best of our knowledge, such methods only
 consist of local solvers complemented by a
 crude global step. The  method introduced below seems to be the first
 one really 
 exhibiting the local/global paradigm in the spirit of  methods used in
 other fields of the engineering sciences.  
 Numerical obervations confirm the
  major practical interest  methodological improvement.

\medskip

Why is  a \emph{linear scaling} plausible for computing $D_\star$? To
justify the fact that the cubic scaling is an estimate by excess of 
the computational task required to solve (\ref{eq:euler}), we argue that
the matrix does not need to be diagonalized. As mentioned above, only the
\emph{orthogonal projector} 
on the subspace generated by the lowest $N$ eigenvectors is to be
determined and \emph{not} the \emph{explicit} values of these lowest $N$ 
eigenvectors. But in order to reach a linear complexity, appropriate
assumptions are necessary, both on
the form of the 
matrices $H$ and $S$, and on the matrix
$D_\star$ solution to 
(\ref{eq:euler}):
\begin{itemize}
\item (H1). The matrices  $H$ and $S$ are assumed sparse, in the sense
  that, for large systems, the number of non-zero coefficients scales as
  $N$. This assumption is not restrictive. In particular, it 
follows from (\ref{eq:defH}) and (\ref{eq:defS}) that it is automatically 
satisfied for Kohn-Sham models as soon as the basis functions are
localized in real space, which is in particular the case for the widely
used atomic orbital basis sets~\cite{cances-handbook};
\item  (H2). A second assumption  is that the matrix $D_\star$ built
  from the solution to (\ref{eq:euler}) is also sparse. This condition
  seems to be fulfilled as soon as the relative gap 
\begin{equation}\label{eq:gam}
\displaystyle \gamma = \frac{\epsilon_{N+1} - \epsilon_{N}}{\epsilon_{N_b}-\epsilon_1}.
\end{equation}
deduced from the solution of (\ref{eq:euler}) is large enough. As
explained in section~\ref{sec:localization} below, this observation can
be supported by qualitative physical arguments. On the other hand, we
are not aware of any mathematical argument of linear algebra that would justify
assumption (H2) in a general setting.
\end{itemize}
We assume (H1)-(H2) in
the following. Current efforts aim at
treating cases when 
the second assumption is not fulfilled, which in particular corresponds to
the case of conducting materials. The problem (\ref{eq:pb}) is then
extremely difficult because the gap
$\gamma$ in (\ref{eq:gam}) being very small, the matrix $D$ is
likely to be dense. Reaching linear complexity is then a challenging
issue, unsatisfactorily solved to date.
 State of the art  linear scaling methods presented in the literature
 experience tremendous
difficulties (to say the least) in such cases. It is therefore
reasonable to  improve in a first step the existing methods in the
setting of assumption (H2), 
before turning to more challenging issues.
                                
\medskip

Before we get to the heart of the matter, we would like to point out
the following feature of the problem under consideration.

\medskip

In practice, Problem (\ref{eq:euler}) has to be
solved \emph{repeatedly}. For instance, it is the inner loop in a nonlinear minimization problem where $H$  depends
self-consistently on $D_\star$. We refer to
\cite{outperform,EDIIS} for 
efficient algorithms to iterate on this nonlinearity and to
\cite{cances-handbook} for
a review on the subject.
Alternatively, or in addition to the above, problem~(\ref{eq:euler}) is
parametrized by the positions of 
the nuclei (both the mean-field operator $H$ and the overlap matrix $S$
indeed depend on these positions), 
and these positions may vary. This is the case 
 in  molecular mechanics (find the optimal configuration 
 of nuclei that gives the lowest possible
 energy to the molecular system), and in
 molecular dynamics as well (the
   positions of nuclei follow the Newton law
 of motion  in the mean-field created by the electrons).
In either case,  problem (\ref{eq:euler}) is
 not be solved \emph{from scratch}.
Because of previous calculations, we
may consider we 
have at our disposal a good initial guess for the solution. The latter
comes from e.g. previous positions of nuclei, or previous
iterations in the outer loop of determination of $H$. In difficult
cases it may even come from a previous computation with a coarse
grained model. In other words, the question addressed reads
{\em solving Problem (\ref{eq:euler}) for some $H+\delta H$ and $S +
  \delta S$ that are small perturbations 
of previous $H$ and $S$ for which the solution is known}. This specific
context allows for a speed up of the algorithm when the initial guess is
sufficiently good. This is the reason why, in the following, we
shall frequently make distinctions between bad and good initial guesses.

\section{Localization in Quantum Chemistry}
\label{sec:localization}
 
\begin{figure}[h]
  \centering
   \epsfig{figure=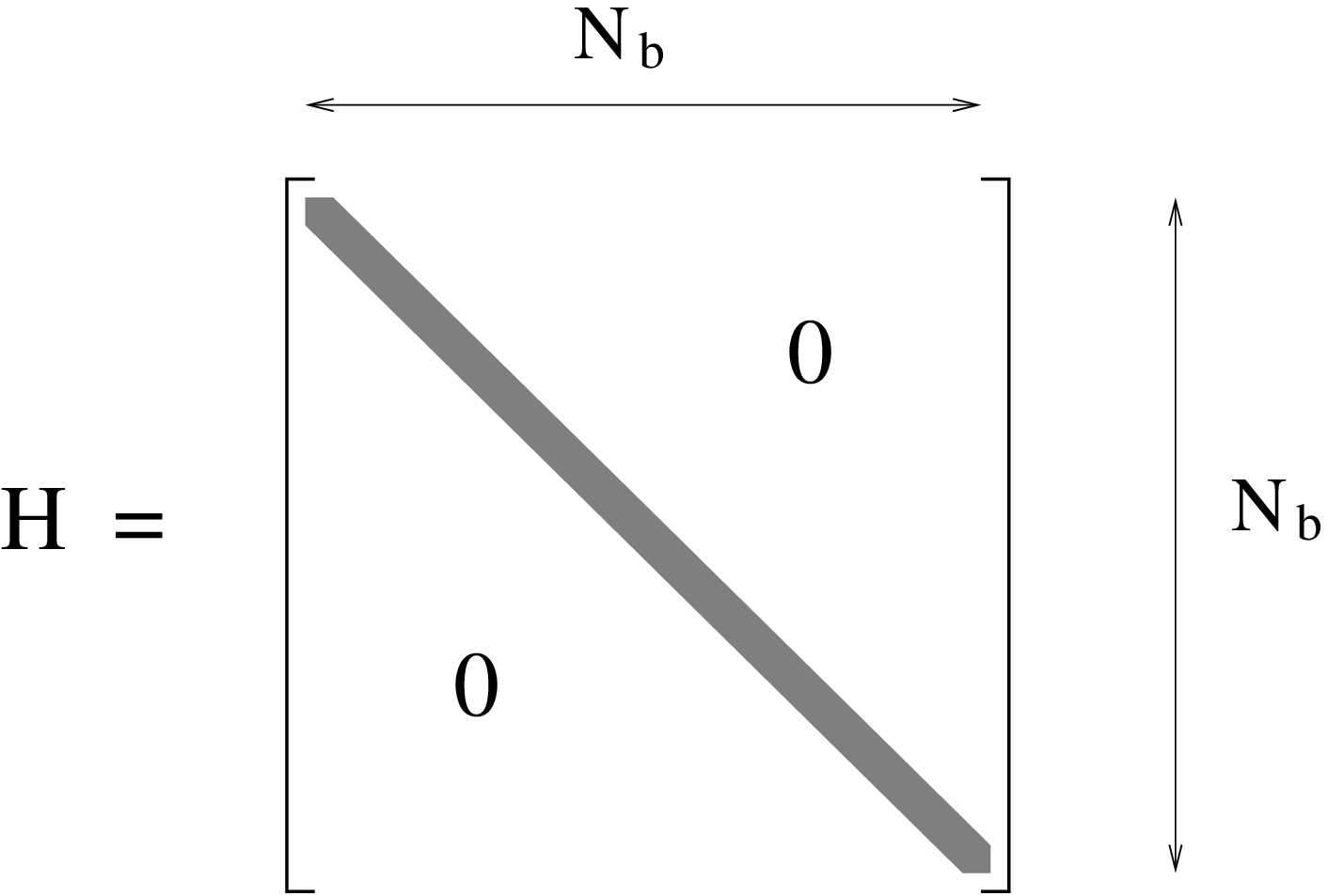,height=5truecm}
     \caption{
     Band structure of the symmetric matrix $H$.
\label{fig:Fband}}
\end{figure}

The
physical system we consider is a long linear molecule (for 
instance a one-dimensional polymer or a nanotube). Let us emphasize that
we do not claim a particular physical relevance of this system. This is
for the purpose of illustration. We believe the
system considered to be a good  representative of a broad class of large
molecular systems that may be encountered practically.
Each atomic orbital $\chi_i$
is centered on one nucleus. Either it is supported in a
ball of small radius \cite{Siesta} (in comparison to the size of the
macromolecule under study), or it has a rapid exponential-like 
or Gaussian-like \cite{gill} fall-off. The  atomic
orbitals are numbered following the orientation of the
molecule. Then, the mean-field
Hamiltonian matrix $H$ whose entries are defined by (\ref{eq:defH})
has the band structure shown in Figure~\ref{fig:Fband}. 

\medskip

\begin{figure}[h]
  \centering
   \epsfig{figure=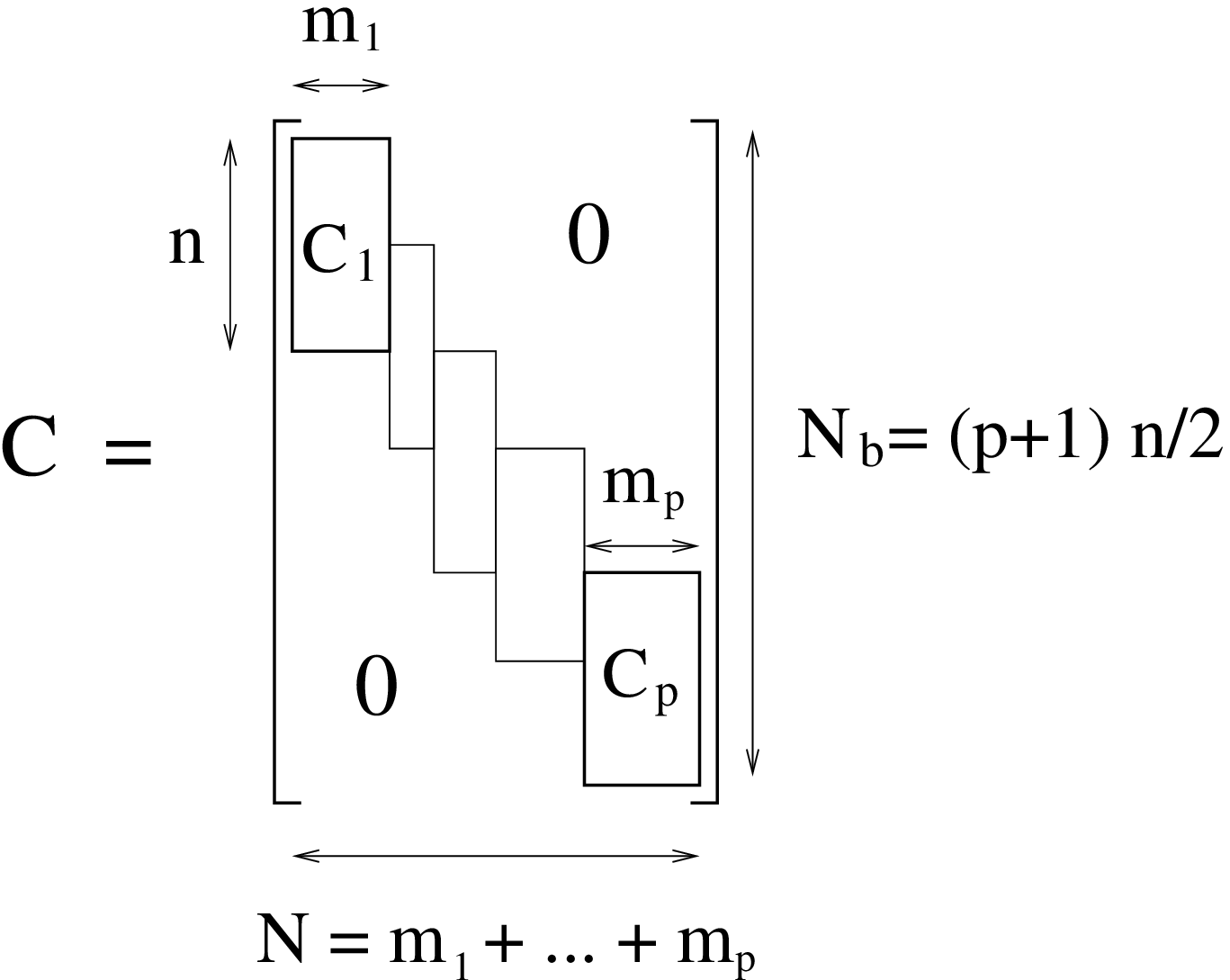,height=5.5truecm}
     \caption{
     Block structure of the  matrices $C$. Note that by
     construction each
     block only overlaps with its nearest
     neighbors.\label{fig:Cblock}}  
\end{figure}

\begin{figure}[h]
  \centering
   \epsfig{figure=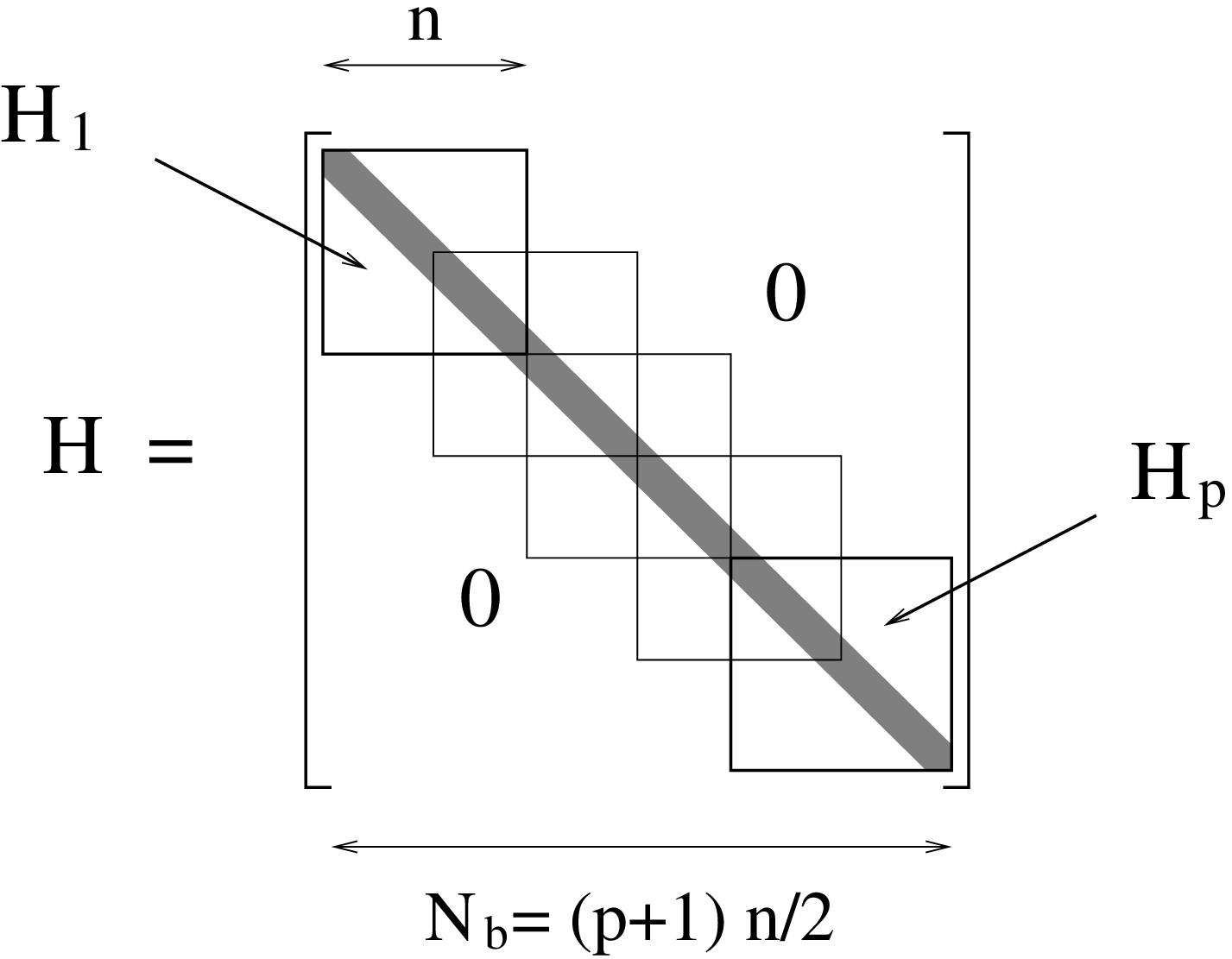,height=5truecm}
     \caption{
     Block structure of the matrix $H$.\label{fig:Fblock}} 
\end{figure}

Although the
eigenvectors of $H$ are {\it a priori} delocalized (most of their
coefficients do not vanish), it seems to be possible to
build a $S$-orthonormal basis of the subspace generated by the lowest $N$
eigenvectors of $H$, consisting of {\it localized} vectors (only a few
consecutive coefficients are non zero). This is motivated by a
physical argument of locality of the interactions \cite{Kohn}. For periodic
systems, the localized vectors correspond to the so-called Wannier
orbitals \cite{Wannier}. It can be proven that in this case, the larger the
band gap, the better the localization of the Wannier orbitals
\cite{Kohn_crystals}. For
insulators, the Wannier orbitals indeed enjoy an exponential fall-off 
rate  proportional to the band gap. For
conductors, the fall-off is only algebraic. As mentioned in the
introduction, we only consider here the former case.
This allows us to assume that there exists some integer $q \ll N_b$, such that 
$N_b/q$ is an integer, for which all of these localized functions can be
essentially expanded on $q$ consecutive atomic orbitals. Denoting by
$n=2q$, we can therefore assume a good
approximation of a solution $C_\star$ to (\ref{eq:infF}) exists, with
the block
structure displayed on Figure~\ref{fig:Cblock}. Note that each
block $C_i$ only overlaps with its nearest neighbors. Correspondingly,
we  introduce the block 
structure of $H$ displayed on Figure~\ref{fig:Fblock}.  The matrix $D$
constructed from a block matrix $C$ 
using (\ref{eq:D}) has the structure represented in Figure~\ref{fig:D}
and satisfies the constraints $D=D^t$, $D^2=D$, Tr($D)$ = $N$. 

\medskip

Let us point out that the integers $q$ and $n=2q$ depend on the
band gap, \emph{not} on the size of the molecule. The condition $n=2q$
is only valid 
for $S=I_{N_b}$. For $S\not=I_{N_b}$, it is replaced by
$n=2q+nbs$   where $2 \, nbs - 1$ is 
the bandwidth of the matrix $S$. 

\medskip

The domain decomposition algorithm we propose aims at searching an
approximate solution to (\ref{eq:infF}) that has the block structure
described above. 

\medskip

For simplicity, we now present our  method
assuming that $S=I_{N_b}$, i.e. that the Galerkin basis $\left\{
  \chi_i \right\}_{1 \le i \le N_b}$ is orthonormal. The
extension of the method to the case when $S\neq I_{N_b}$ is
straightforward. Problem (\ref{eq:infF}) then reads  
\begin{equation}
  \label{eq:infF2}
\inf \biggl\{\mbox{\textrm{Tr}}\Big(HCC^{t}\Big), \quad C\in
\mm^{N_{b},N}(\RR), \; C^{t}C=I_{N} \biggr\}.
\end{equation}

\medskip

Our approach consists in solving an approximation of problem
(\ref{eq:infF2}) obtained by minimizing the exact energy 
$\dps \mbox{\textrm{Tr}}\Big(HCC^{t}\Big)$ on the set of the matrices
$C$ which have the block structure displayed on Figure~\ref{fig:Cblock}
and satisfy the constraint $C^{t}C=I_{N}$. The resulting minimization
problem can be recast as
\begin{eqnarray}
\mbox{\textrm{inf}}\bigg\{
\sum_{i=1}^{p}\mbox{\textrm{Tr}}\left(H_iC_iC_i^{t}\right), & &  C_i\in
\mm^{n,m_i}(\RR),\quad m_i \in \NN, \quad C_i^{t}C_i=I_{m_i} \quad \forall \; 1\leq i\leq p, 
\nonumber \\
& & C_{i}^{t}TC_{i+1}=0 \quad \forall \; 1\leq i\leq p-1, \quad \sum_{i=1}^p m_i = N  \bigg\}.
\label{eq:infF-blocks}
\end{eqnarray}
In the above formula, $T \in {\cal M}^{n,n}(\RR)$  is the matrix defined by
\begin{equation}
\label{eq:matrices_t}
T_{kl} = 
\left\{ \begin{array}{ll} 
1 & \mbox{ if }  k - l = q \\
0 & \mbox{ otherwise}
\end{array} \right.
\end{equation}
and $H_i \in {\cal M}^{n,n}(\RR)$ is a symmetric submatrix of $H$ (see
Figure~\ref{fig:Fblock}). Indeed, 

\begin{figure}[h]
  \centering
   \epsfig{figure=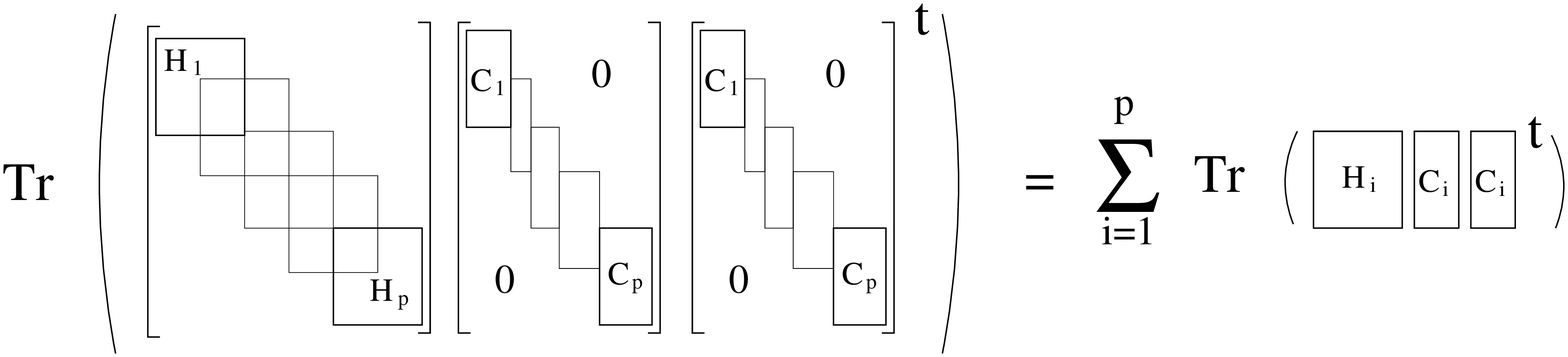,height=3truecm}
\end{figure}

\noindent
and

\begin{figure}[h]
  \centering
   \epsfig{figure=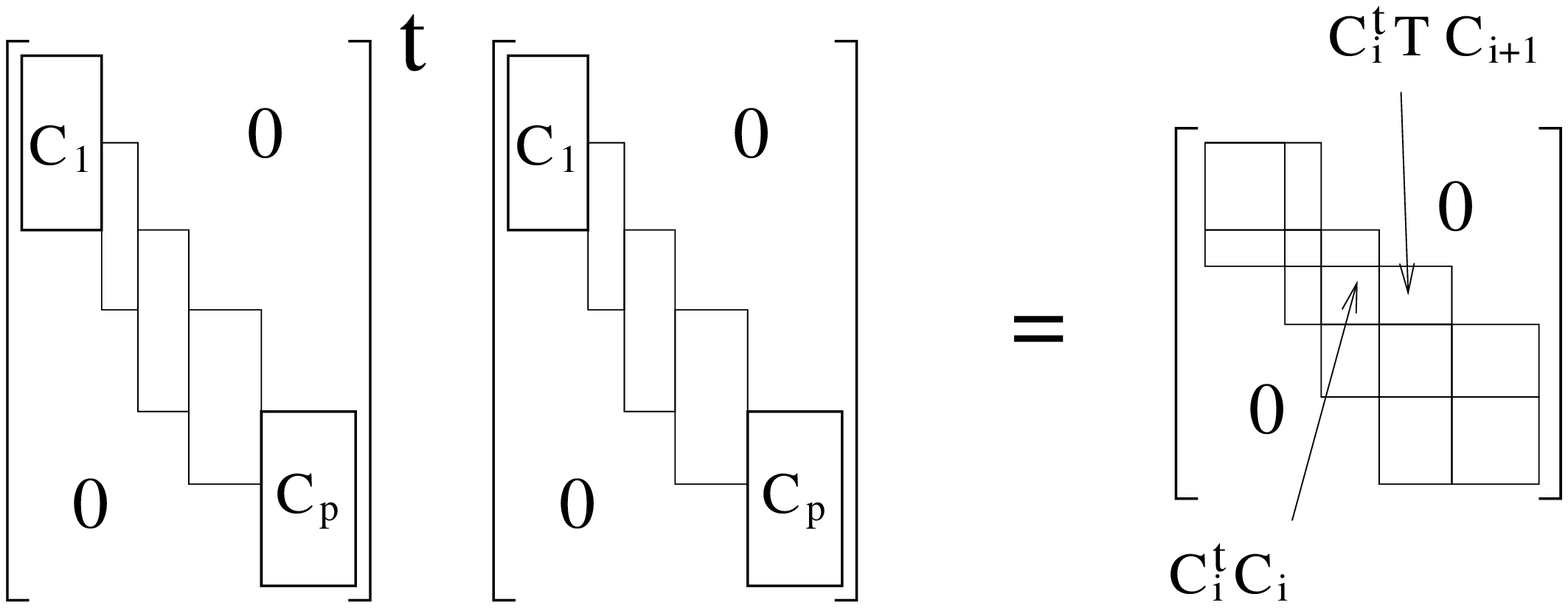,height=3truecm}
\end{figure}

In this way, we replace the $\frac{N(N+1)}2$ {\em global} scalar
constraints $C^tC=I_N$ involving vectors of size $N_b$, by the 
$\sum_{i=1}^p \frac{m_i (m_i+1)}2$ {\em local} scalar constraints
$C_i^t C_{i}=I_{m_i}$ and the $\sum_{i=1}^{p-1} m_i m_{i+1}$
{\em local} scalar constraints $C_i^t T
C_{i+1}=0$, involving 
vectors of size $n$. We would like to emphasize that we can
obtain in this way a basis of the vector space generated by the lowest
$N$ eigenvectors of $H$, but not the eigenvectors themselves. This
method is therefore not directly applicable to standard diagonalization
problems. 

\medskip

Our algorithm searches for the solution to (\ref{eq:infF-blocks}), not
to (\ref{eq:infF2}). More rigorously stated, we search for the
solution to the  Euler-Lagrange equations of (\ref{eq:infF-blocks}):
\begin{equation}
\label{eq:euler-blocks}
 \left\{
\begin{array}{ccll}
\displaystyle H_iC_i &=& C_iE_i + T^t C_{i-1}\Lambda_{i-1,i} +
T C_{i+1}\Lambda_{i,i+1}^t & \qquad 1 \le i \le p,\\ 
\displaystyle  C_i^{t}C_i &=&I_{m_i} & \qquad  1 \le i \le p,\\
\displaystyle  C_{i}^{t}T C_{i+1} &=&0&  \qquad 1 \le i \le p-1,
\end{array}\right.
\end{equation}
where by convention 
\begin{equation}
\label{eq:convention}
 C_0 = C_{p+1} = 0.
\end{equation}
The matrices $(E_i)_{1 \le i \le p}$ and $(\Lambda_{i,i+1})_{1
  \le i \le p-1}$ respectively 
denote the matrices of Lagrange multipliers associated with the
  orthonormality constraints $C_i^{t}C_i=I_{m_i}$ and $C_{i}^{t}T
  C_{i+1} = 0$. The $m_i \times m_i$ matrix $E_i$ is symmetric. The matrix 
$\Lambda_{i,i+1}$ is of size $m_i \times m_{i+1}$.
The above equations can be easily derived by considering the Lagrangian
\begin{eqnarray*}
{\cal L}\left(\left\{ C_i \right\}, \left\{ E_i \right\}, \left\{
  \Lambda_{i,i+1} \right\} \right) & = & \sum_{i=1}^p \tr\left( H_iC_iC_i^t
\right) + \sum_{i=1}^p \tr \left( \left(C_i^tC_i-I_{m_i}\right) E_i
\right) \\ & & +   \sum_{i=1}^{p-1}  \tr \left( C_i^tTC_{i+1}\Lambda_{i,i+1}^t
\right).
\end{eqnarray*}

The block structure imposed on the matrices
clearly lowers the dimension of the search space we have to
explore.
However, this simplification comes at a  price. First,
problem~(\ref{eq:infF-blocks}) only {\em approximates} problem
(\ref{eq:infF2}). Second, (\ref{eq:infF-blocks}) may have local, non global,
minimizers, whereas all the local minimizers of 
(\ref{eq:infF2}) are global. There are thus {\it a priori} many spurious 
solutions of the Euler Lagrange equations (\ref{eq:euler-blocks})
associated with~(\ref{eq:infF-blocks}).

\medskip

A point is that the sizes $(m_i)_{1 \le i \le p}$ are  not
{\it a priori} prescribed. In our approach, they  are ajusted during the
iterations.  
We shall see how in the sequel.

\section{Description of the domain decomposition algorithm}

\subsection{Description of a simplified form}
\label{sec:simdesc}

For pedagogic purpose, we first consider the following  problem
\begin{equation}  \label{eq:inf-vector-biblock}
    \inf \left\{ \langle H_1Z_1,Z_1\rangle+\langle H_2Z_2,Z_2\rangle,
    \quad Z_i \in{\RR}^{N_b}, \; 
\langle Z_i,Z_i\rangle = 1, \; \langle
    Z_1,Z_2\rangle=0 \right\}.
\end{equation}
Problem (\ref{eq:inf-vector-biblock}) is a particular occurence of
(\ref{eq:infF-blocks}).
We have denoted by
$\langle\cdot,\cdot\rangle$ the 
standard Euclidean scalar product on ${\RR}^{N_b}$.

For  (\ref{eq:inf-vector-biblock}), the algorithm is defined in the
following simplified form. Choose $(Z_1^0,Z_2^0)$ satisfying the
constraints and construct the sequence $(Z_1^k,Z_2^k)_{k \in \NN}$ by
the following iteration procedure. Assume $(Z_1^k,Z_2^k)$ is
known, then 
\begin{itemize}
\item{ Local step}. Solve 
\begin{equation}
  \label{eq:s11}
  \left\{\begin{array}{lll}
  \widetilde Z_1^k = \mbox{\textrm{arginf}}\big\{\langle H_1Z_1,Z_1\rangle,
    \,\,Z_1\in{\RR}^{N_b},\,\,\,\langle Z_1,Z_1\rangle=1\,\,\langle Z_1,Z_2^k\rangle=0\big\},\\
\widetilde Z_2^k = \mbox{\textrm{arginf}}\big\{\langle H_2Z_2,Z_2\rangle,
    \,\,Z_2\in{\RR}^{N_b},\,\,\,\langle Z_2,Z_2\rangle=1\,\,\langle
  \widetilde Z_1^k ,Z_2\rangle=0\big\} \; ;
  \end{array}
\right.
\end{equation}

\item{ Global step}. Solve
  \begin{equation}
  \label{eq:s21} 
  \alpha^\ast= \mbox{\textrm{arginf}} \big\{\langle
  H_1Z_1,Z_1\rangle+\langle H_2Z_2,Z_2\rangle, \; \alpha\in{\RR}\big\} 
  \end{equation}
where
  \begin{equation}
  \label{eq:s21bis}
  Z_1=\frac{\widetilde Z_1^k+\alpha\widetilde
  Z_2^k}{\sqrt{1+\alpha^2}},\,\,Z_2=\frac{-\alpha\widetilde Z_1^k+\widetilde
  Z_2^k}{\sqrt{1+\alpha^2}},
  \end{equation}
and set
\begin{equation}
\label{eq:newit}
  Z_1^{k+1}=\frac{\widetilde Z_1^k+\alpha^\ast\widetilde
  Z_2^k}{\sqrt{1+(\alpha^\ast)^2}},\quad 
 Z_2^{k+1}=\frac{-\alpha^\ast\widetilde Z_1^k+\widetilde Z_2^k}
{\sqrt{1+(\alpha^\ast)^2}}.
\end{equation}
\end{itemize}

\medskip

In the $k$-th iteration of the local step, we first fix $Z_2 = Z_2^k$
   and optimize over $Z_1$ to obtain $\tilde{Z}_1^k$. Then we fix $Z_1 =
   \tilde{Z}_1^k$ and optimize over $Z_2$ to obtain $\tilde{Z}_2^k$.
   This local step monotonically reduces the
   objective function, however, it may not converge to
   the global optimum. The technical problem is that the Lagrange multipliers
   associated with the constraint $\langle Z_1,Z_2\rangle = 0$ may
   converge to different 
   values in the two subproblems associated with the local step.
   In the global step, we optimize the {\em sum} $\langle H_1
   Z_1,Z_1\rangle + \langle H_2 Z_2,Z_2\rangle$ over the subspace spanned by
   $\tilde{Z}_1^k$ and $\tilde{Z}_2^k$, subject to the constraints
   in~(\ref{eq:inf-vector-biblock}). 
   The global step again reduces the value of the objective function since
   $\tilde{Z}_1^k$ and $\tilde{Z}_2^k$ are feasible in the global step.
   It can be shown that the combined algorithm (local step + global step)
   monotonically decreases the objective function and globally
   converges to an optimal solution of~(\ref{eq:inf-vector-biblock}).

   This algorithm operates at two levels: a fine level where we
   solve two problems of dimension $N_b$ rather than one problem of
   dimension $2N_b$; a coarse level where we solve a problem of dimension~$2$.
   Left by itself, the fine step converges to a suboptimal
   solution of~(\ref{eq:inf-vector-biblock}). Combining the fine step with
   the global step yields convergence to a global optimum.

In addition to providing a pedagogic view on the general algorithm
presented in the following section, the simplified form
(\ref{eq:s11})-(\ref{eq:newit}) has a theoretical interest. In contrast
to the general algorithm for which we cannot provide
a convergence analysis, the simplified form
(\ref{eq:s11})-(\ref{eq:newit})  may be analyzed mathematically, at least
in the particular situation when 
$H_1=H_2=H$. Then solving (\ref{eq:inf-vector-biblock}) amounts to
searching for the lowest two eigenelements of the matrix $H$. 
Notice that the global step (\ref{eq:s21})-(\ref{eq:newit})  is then
unnecessary because the 
functional to minimize in (\ref{eq:s21})  does not depend on $\alpha$. 

However, we can show that the iterations (\ref{eq:s11}) converge
in the following sense.  The 2-dimensional vector space spanned by the
lowest two eigenvalues of $H$ is reached asymptotically. This occurs
 under an appropriate condition on the matrix $H$. The latter is  a
 condition of separation of  the eigenvalues, namely 
$\displaystyle\epsilon_2-\epsilon_1 < \epsilon_3 - \epsilon_2$ with obvious
notation. The gap $\epsilon_3-\epsilon_2$ gives the speed of
convergence. For brevity, we do not detail the proof here
(see~\cite{theseMax}). Future work on the numerical analysis of
more general cases is in progress.

\subsection{Description of the algorithm} 
\label{sec:gendesc}

We define, for all $p$-tuple $(C_i)_{1\leq i\leq p}$,
\begin{equation}
\displaystyle \me\Big((C_i)_{1\leq i\leq p}\Big) =
\sum_{i=1}^p\mbox{\textrm{Tr}}\Big(H_iC_iC_i^t\Big), 
\end{equation}
and set by convention
\begin{equation} 
 U_{0} = U_{p} = 0.
\end{equation}
We introduce an integer $\epsilon$, initialized to one, that will
alternate between the values zero and one during the iterations.

\medskip

At iteration $k$, we have at hand a set of block sizes
$(m_i^k)_{1 \le i \le p}$ and a set of matrices $(C_i^k)_{1 \le i \le
  p}$ such that $C_i^k \in {\cal M}^{n,m_i^k}(\RR)$,
$[C_i^k]^tC_i^k=I_{m_i^k}$, $[C_i^k]^t T C_{i+1}^k = 0$. We now explain 
how to compute the
new iterate $(m_i^{k+1})_{1 \le i \le p}$,  $(C_i^{k+1})_{1 \le i \le
  p}$.

\medskip

\noindent
{\bf Multilevel Domain Decomposition (MDD) algorithm}

\medskip

\begin{enumerate}
\item[] {\bf $\bullet$ Step 1: Local fine solver}. 

\begin{enumerate}
\item For each $i$, diagonalize the matrix $H_{2i+\epsilon}$ in the
  subspace  
$$
V_{2i+\epsilon}^k = \left\{ x \in \RR^{n}, \quad 
\left[C_{2 i + \epsilon - 1}^k\right]^t T x = 0, \quad x^t T C_{2i + \epsilon + 1}^k
 = 0 \right\}, 
$$
i.e. diagonalize $P_{2i+\epsilon}^k H_{2i+\epsilon} P_{2i+\epsilon}^k$
where $P_{2i+\epsilon}^k$ is the orthogonal projector on $V_{2i+\epsilon}^k$.
This provides (at least) $n - m_{2i +\epsilon - 1}^k - m_{2i +\epsilon+
  1}^k$ real eigenvalues 
$\lambda_{2i+\epsilon,1}^k \le \lambda_{2i+\epsilon,2}^k \le \cdots $
and associated orthonormal vectors $x_{2i+\epsilon,j}^k$. The
latter are $T$-orthogonal to the column vectors of $C_{i-1}^k$ and
$C_{i+1}^k$.  

\item Sort  the eigenvalues $(\lambda_{2i+\epsilon,j}^k)_{i,j}$  in
  increasing order, and select the  lowest $\dps \sum_i 
  m_{2i+\epsilon}$ of them. For each $i$, collect in
  block $\#2i+\epsilon$  
  the eigenvalues $\lambda^k_{2i+\epsilon,j}$  selected. New
  intermediate block sizes $\bar m_{2i+\epsilon}^{k}$ are defined. 

\item For each $i$, collect the lowest $\bar m_{2i+\epsilon}^{k}$ vectors
  $x_{2i+\epsilon,j}^k$  in the $n \times
  \bar m_{2i+\epsilon}^{k}$ matrix $\overline C_{2i+\epsilon}^{k}$.

\item For each $i$, diagonalize the matrix $H_{2i+\epsilon+1}$ in the
  subspace  
$$
V_{2i+\epsilon+1}^k = \left\{ x \in \RR^{n}, \quad 
 \left[\overline C_{2i + \epsilon}^k\right]^t T x = 0, \quad x^t T
 \overline C_{2i + \epsilon +2}^k = 0 \right\}
$$
in order to get eigenvalues $\lambda_{2i+\epsilon+1,1}^k \le
\lambda_{2i+\epsilon+1,2}^k \le \cdots $ 
and associated orthonormal vectors $x_{2i+\epsilon+1,j}^k$. The
latter are $T$-orthogonal to the column vectors of $ \overline C_{2i +
  \epsilon}^k$ and $\overline C_{2i + \epsilon +2}^k$.  

\item  Sort all the  eigenvalues 
$\displaystyle\bigl\{(\lambda_{2i+\epsilon+1,j}^k)_{i,j},
(\lambda_{2i+\epsilon,j}^k)_{i,j}\bigr\}$  
in
  increasing order. Select the lowest $N$.   For each $l$, collect in
  block $\#l$   the
  eigenvalues $\lambda^k_{l,j}$ selected. 
 New intermediate
  block sizes $(m_{l}^{k+1})_{1 \le l \le p}$ are thus defined.  

\item Set $\widetilde C_l^{k} = \dps \left[
    x_{l,1}^k | \cdots |x_{l,m_{l}^{k+1}}^k 
    \right]$.

\item Replace $\epsilon$ by $1-\epsilon$ and proceed to step~2 below.

\end{enumerate}

\item[] {\bf $\bullet$ Step 2: global coarse solver}. Solve
\begin{equation}
\label{eq:global-manyblock}
\displaystyle \muu^\ast =  \mbox{\textrm{arginf}}\Big\{f(\muu),\; \muu =
(U_{i})_i,\;\forall 1 \le i
\le p-1 \;\; U_i \in
\mm^{m_{i+1},m_{i}}(\RR)\Big\},
\end{equation}
where
\begin{equation}
\label{eq:global-manyblock1}
\displaystyle f(\muu) = \me\bigg(\Big(C_i(\muu)\big(C_i(\muu)^tC_i(\muu)\big)^{-\frac{1}{2}}\Big)_i\bigg),
\end{equation}
and
\begin{equation}
\label{eq:global-manyblock2}
C_i(\muu)    =  \widetilde{C}_i^k + T
\widetilde{C}_{i+1}^k U_i\Big( [\widetilde{C}_{i}^k]^t T T^t
\widetilde{C}_{i}^k \Big) - T^t
\widetilde{C}_{i-1}^k U_{i-1}^t\Big([\widetilde{C}_{i}^k]^tT^t T
\widetilde{C}_{i}^k\Big). 
\end{equation}
Next set, for all $1 \le i \le p$,
\begin{equation} \label{eq:defCkp1}
C_i^{k+1} = C_i \big(\muu^\ast\big) \; \left( C_i \big(\muu^\ast\big)^t
  \,  C_i \big(\muu^\ast\big) \right)^{-1/2} \; .
\end{equation}
Note that $\dps \left[ C_i^{k+1} \right]^t T C_{i+1}^{k+1}=0$ (this
follows from $T^2=0$).
\end{enumerate}

We think of the even indexed unknowns $C_{2i}$ as the black variables
   and the odd indexed unknowns $C_{2i+1}$ as the white variables.
   In the first phase of the local fine solver, we optimize over
   the white variables while holding the black variables fixed. In
   the second phase of the local fine solver, we optimize over the
   black variables while holding the white variables fixed. In the
   global step, we perturb each variable by a linear combination of
   the adjacent variables. The matrices ${\muu} = (U_{i})_i$ in
   (\ref{eq:global-manyblock}) play the same role as the real parameter
   $\alpha$ in (\ref{eq:s21}). The perturbation is designed so that the
   constraints are satisfied. The optimization is performed over the matrices
   generating the linear combinations. In the next iteration, we interchange
   the order of the optimizations: first optimize over the black variables
   while holding the white variables fixed, then optimize over the white
   variables while holding the black variables fixed.

\medskip
 
\begin{figure}[h]
  \centering
   \epsfig{figure=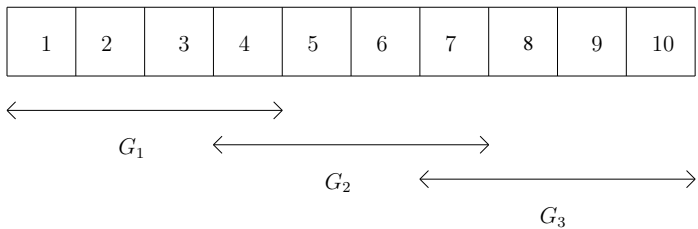,height=3truecm}
    \caption{
     Collection of $p=10$ blocks into $r=3$ groups.\label{fig:groupe}}
\end{figure}

Let us point out that an accurate solution to
(\ref{eq:global-manyblock}) is not needed.  
In practice, we reduce the computational cost of the global step, by
using again a domain decomposition method. The blocks $(C_i)_{1 \le 
  i \le p}$ are collected in $r$ overlapping groups $(G_l)_{1 \le l \le
  r}$ as shown in Figure~\ref{fig:groupe}. Problem
(\ref{eq:global-manyblock}) is solved first for 
the blocks  $(G_{2l+1})$, next for the blocks $(G_{2l})$. Possibly,
this procedure is repeated a few times. The advantage of this strategy
is that the
computational time of the global step scales linearly with $N$.
In addition, it is parallel in nature. 
The solution of~(\ref{eq:global-manyblock}) for a given group is
performed by a few steps of a
Newton-type algorithm. Other preconditioned iterative methods 
could also be considered.

\subsection{Comments on the local step}

The local step is based on a checkerboard iteration technique. 

\medskip

When $\epsilon=1$, steps 1a-1c search for a solution 
$(\bar m_{2i+1}^{k},\overline C_{2i+1}^{k})_i$ to the problem
\begin{eqnarray*} 
\mbox{\textrm{inf}}\bigg\{
\sum_{i} \mbox{\textrm{Tr}}\left(H_{2i+1}C_{2i+1}C_{2i+1}^{t}\right),
& & C_{2i+1} \in 
\mm^{n,m_{2i+1}}(\RR),\quad C_{2i+1}^{t}C_{2i+1}=I_{m_{2i+1}}, \\
& & 
[C_{2i}^k]^{t}TC_{2i+1}=0,  \quad C_{2i+1}^{t}TC_{2i+2}^k=0,  \\
& & 
 m_{2i+1} \in \NN, \quad \sum_{i} m_{2i+1} = 
\sum_{i} m_{2i+1}^k  \bigg\} .
\end{eqnarray*}
During steps 1a-1c, the ``white'' blocks $C_{2i}^k$ are kept
fixed. The ``black'' blocks $C_{2i+1}^k$ are optimized
under the orthogonality constraints imposed by the ``white''
blocks. A point is that most of the computational effort can
be done {\em in parallel}. Indeed, for $p$ even, say, performing step 1a amounts to solving
$p/2$  {\em independent} diagonalisation problems of size~$n$.

\medskip

Likewise, steps 1d-1f solve 
\begin{eqnarray*} 
\mbox{\textrm{inf}}\bigg\{
\sum_{i=1}^p \mbox{\textrm{Tr}}\left(H_{i}C_{i}C_{i}^{t}\right),
& & C_{i} \in 
\mm^{n,m_{i}}(\RR),\quad C_{i}^{t}C_{i}=I_{m_{i}},  \quad  m_{i} \in
\NN, \quad \sum_{i} m_{i} = N
\\ &  &
[\overline
C_{2j-1}^k]^{t}TC_{2j}=0 , \quad
[C_{2j}]^{t}T [\overline C_{2j+1}]^k=0, \\ &  &
  0 \le m_{2j+1} \le \bar m_{2j+1}^k, \quad C_{2j+1} \subset \overline
  C_{2j+1}^k  
  \bigg\},
\end{eqnarray*}
where the notation $C_{2j+1} \subset \overline
  C_{2j+1}^k$ means that each column of $C_{2j+1}$ is a column of 
 $\overline C_{2j+1}^k$. Here again, most of the computational effort
  can be performed in parallel. 

\medskip

When $\epsilon=1$, ``black'' vectors (i.e. vectors belonging to blocks with
odd indices) are allowed to become ``white'' vectors, but the reverse is
forbidden. In order to symmetrize the process, $\epsilon$ is replaced by
 $1-\epsilon$ in the next iteration.

\medskip

We wish to emphasize that, although called {\it local}, this step
already accounts for some global concern. Indeed, and it is a key point
of the local step, substeps (b) and (e) sort the \emph{complete} set of
eigenvalues generated locally. This, together with the update of the size
$m_i$ of the blocks, allows for a preliminary propagation of the
information throughout the whole system. The global step will complement
this.

\medskip

Finally, let us mention that in the local steps, (approximate)
$T$-orthogonality is obtained by a Householder orthonormalization
process. The required orthonormality criterion is
\begin{equation} \label{eq:epsilonL}
\forall \; 1 \le i \le p-1, \quad
\big\|[\tilde{C}^k_i]^tT\tilde{C}^k_{i+1} \big\| \leq \epsilon_L, 
\end{equation}
where $\epsilon_L > 0$ is a threshold to be chosen by the user.

\subsection{Comments on the global step} \label{sec:global}

Let us briefly illustrate the role played by the global step. For
simplicity, we consider the case of two blocks of same initial size
$m_1=m_2=m$ and we assume that $m_1$ and $m_2$ do not vary during the
iterations. If only the local step is performed, 
then the new iterate
$$(C^{k+1}_1,C_2^{k+1})=(\widetilde{C}_1^{k},\widetilde{C}_2^{k})$$
does not necessarily satisfies  (\ref{eq:euler-blocks}). Indeed, there is no
reason why the 
Lagrange multipliers corresponding to the two contraints $C^t
T C_2^k = 0$ (step 1a when $\epsilon=1$) on the one hand 
and $[\widetilde{C}_1^k]^t T C = 0$  (step 1d when $\epsilon=1$) 
 on the other hand should be the same. The
 global step
  \emph{asymptotically} enforces the  equality of Lagrange
  multipliers. This  is a way to
 account for a global feature of the problem. 

Let us emphasize this
 specific point. Assume  
 $U^\ast = 0$ in the global step of the  $k$-th iteration of the
 algorithm, or in other words that the global step is not effective at
 the $k$-th iteration. Then it implies that the output
$(\widetilde{C}_1, \widetilde{C}_2) = (\widetilde{C}_1^k,
 \widetilde{C}_2^k)$ of the local step already satisfies
 (\ref{eq:euler-blocks}).
Indeed,
\begin{equation}
\displaystyle f(U) = \mbox{\textrm{Tr}}\Big(J_1(U)C_1(U)^tH_1C_1(U)\Big) + \mbox{\textrm{Tr}}\Big(J_2(U)C_2(U)^tH_2C_2(U)\Big)
\end{equation}
with $J_i(U) = \Big(C_i(U)^tC_i(U)\Big)^{-1}$ for $i=1,2$. Since
\begin{eqnarray*}
\Big(J_1(U)\Big)^{-1} &=& I_m +
\Big(\widetilde{C}_1^tTT^t\widetilde{C}_1\Big) U^t 
\Big(\widetilde{C}_2^tT^tT\widetilde{C}_2\Big) U
\Big(\widetilde{C}_1^tTT^t\widetilde{C}_1\Big),\\
\Big(J_2(U)\Big)^{-1} &=& I_m + 
\Big(\widetilde{C}_2^tT^tT\widetilde{C}_2\Big) U
\Big(\widetilde{C}_1^tTT^t\widetilde{C}_1\Big) U^t
\Big(\widetilde{C}_2^tT^tT\widetilde{C}_2\Big),
\end{eqnarray*}
we have  $\nabla J_1(0) = \nabla J_2(0) = 0$. The matrix $U$ being a square
matrix of dimension $m$, for all $1 \le i,j \le m$,
\begin{eqnarray}
\displaystyle \frac{1}{2}\frac{\partial f}{\partial U_{ij}}(0) &=&
\displaystyle 
\mbox{\textrm{Tr}}\bigg( \left[\frac{\partial C_1}{\partial
  U_{ij}}(0) \right]^tH_1\widetilde{C}_1\bigg) +
\mbox{\textrm{Tr}}\bigg(  \left[\frac{\partial C_2}{\partial
  U_{ij}}(0) \right]^tH_2\widetilde{C}_2\bigg) \nonumber\\
&=& \displaystyle
\bigg( \Big(\widetilde{C}_1^tTT^t\widetilde{C}_1\Big)
\widetilde{C}_1^t H_1 T \widetilde{C}_2
\bigg)_{ji} -
\bigg(\widetilde{C}_1^tTH_2\widetilde{C}_2
\Big(\widetilde{C}_2^tT^tT\widetilde{C}_2\Big)\bigg)_{ji}
\nonumber\\ 
&=& \displaystyle \bigg(\Big(\widetilde{C}_1^tTT^t\widetilde{C}_1\Big)(\Lambda_{1}-\Lambda_{2})\Big(\widetilde{C}_2^tT^tT\widetilde{C}_2\Big)\bigg)_{ji},
\end{eqnarray}
where $\Lambda_1$ and $\Lambda_2$ are defined by
\begin{equation}
\left\{
\begin{array}{ccc}
\displaystyle H_1\widetilde{C}_1 = \widetilde{C}_1E_1 + T\widetilde{C}_{2}\Lambda_{1}^t,\\
\displaystyle H_2\widetilde{C}_2 = \widetilde{C}_2E_2 +
T^t\widetilde{C}_{1}\Lambda_{2}.\\ 
\end{array}\right.
\end{equation}
As $U^\ast = 0$ implies
\begin{equation}
\displaystyle \forall \; 1 \leq i,j \leq  m \quad  \frac{\partial f}{\partial
  U_{ij}} (0) = 0,
\end{equation}
we conclude that  $ \Lambda_1 = \Lambda_2$ if the matrices
$\Big(\widetilde{C}_1^tTT^t\widetilde{C}_1\Big)$ and
$\Big(\widetilde{C}_2^tT^tT\widetilde{C}_2\Big)$ are invertible, which  is
generally the case when $n \gg 2m$. Consequently,  
(\ref{eq:euler-blocks})  is satisfied by
$(\widetilde{C}_1, \widetilde{C}_2)$.

On the other hand,  when $n$ is not much larger that $2m$, the above
matrices are not 
invertible and (\ref{eq:euler-blocks}) is usually not satisfied. In this
case,  the global step is  slightly modified in order 
to recover (\ref{eq:euler-blocks}) and thus improve the efficiency of
the global step. We replace
(\ref{eq:global-manyblock2})  by  
\begin{equation}
\label{eq:global-manyblock2-modified}
\forall \; 1 \le i \le p, \quad C_i(\muu)    =  \tilde{C}_i^k + T
\widehat{C}_{i+1}^k U_i\Big( [\widehat{C}_{i}^k]^t T T^t
\widehat{C}_{i}^k \Big) - T^t
\widehat{C}_{i-1}^k U_{i-1}^t\Big([\widehat{C}_{i}^k]^tT^t T
\widehat{C}_{i}^k\Big) 
\end{equation} 
where $\widehat{C}_i^k$ is a block formed by  vectors collected in the
vector space defined by  $\widetilde{C}_i^k$. These vectors are selected  using a modified
Gram-Schmidt orthonormalization process. The size of the blocks
$\widehat{C}_i^k$ is appropriately chosen.
The larger the blocks $\widehat{C}_i^k$, the more precise
 the global step but the worse the conditioning of  the optimization problem.
In addition, since the
global step is the most demanding step of the algorithm, considerations
both on the 
 computational time and in terms of memory are accounted for when fixing
 the sizes  of the blocks $\widehat{C}_i^k$.

Our numerical experiments show that when the global step is performed (using (\ref{eq:global-manyblock2}) or
(\ref{eq:global-manyblock2-modified}), depending on $n$ and $m$), 
the blocks $(C^{k+1}_i)_i$ do not exactly satisfy the orthonormality
constraint, 
owing to evident round-off errors. All the linear scaling algorithms
have difficulties in ensuring this constraint and our MDD approach is no
exception. The tests performed
however show that the constraint
remains satisfied throughout the iterations within a good degree of accuracy.

\section{Numerical tests}
\label{sec:numerical}

An extensive set of numerical tests was performed to illustrate the
important features of the domain decomposition algorithm introduced
above, and to compare it with a standard scheme, commonly
used in large scale electronic structure calculations.  

\subsection{Setting of the algorithm and of the tests}

\paragraph{Molecular systems used for the tests}

Numerical tests on the algorithm presented above were performed on three
chemical systems. The first two systems both have formula
COH-(CO)$_{n_m}$-COH. They  differ in their Carbon-Carbon interatomic
distances. For system  $\mathcal{P}_1$, this distance is fixed to~5
atomic units, while it is fixed to~4 for system
$\mathcal{P}_2$. On the other hand, our third system, denoted by
$\mathcal{P}_3$ has formula CH$_3$-(CH$_2$)$_{n_m}$-CH$_3$.

For each of the three systems $\mathcal{P}_1$, $\mathcal{P}_2$,
$\mathcal{P}_3$, several numbers $n_m$ of monomers were considered. A
geometry optimization was performed using the GAUSSIAN
package~\cite{GAUSSIAN98} in order to fix the internal geometrical
parameters of the 
system. The only exception to this is the Carbon-Carbon distance for
$\mathcal{P}_1$ and $\mathcal{P}_2$, 
which, as said above,  is fixed {\it a priori}. Imposing the
Carbon-Carbon distance allows to control the sparsity of the matrices
$H$ and $S$ (the larger the distance, the sparser the
matrices). Although not physically relevant, fixing the Carbon-Carbon
distance is therefore useful for the purpose of numerical tests.

\medskip

\paragraph{Data, parameters and initialization} For an extremely large
number $n_m$ of monomers, the matrices $H$, $S$, 
and $D_\star$ cannot be generated directly with the GAUSSIAN package. We
therefore  
make a periodicity assumption. For large values of $n_m$, these matrices
approach a periodic 
pattern (leaving apart, of course, the ``boundary layer'', that is the terms
involving orbitals close to one end of the linear molecule). So, we
first fix some  $n_m$ sufficiently large, but for which a direct
calculation with Gaussian is feasible, and construct $H$, $S$. 
The matrices $H$ and $S$, as well as the ground-state density matrix
$D_\star$, and the ground-state energy $E_0$, are then obtained 
for arbitrary large $n_m$ assuming periodicity out of the ``boundary layer''.
Likewise, the
gap $\gamma$ in the eigenvalues of $H$ is observed to be constant, for
each system, irrespective of 
the number $n_m$ of polymers, supposedly large. Proceeding so, the gap
for systems  $\mathcal{P}_1$, $\mathcal{P}_2$,
and $\mathcal{P}_3$ is respectively evaluated to  0.00104,   0.00357,
and   0.0281.
 
For our MDD approach,  localization parameters are needed. They are shown in
Table~\ref{table} below. Additionally, we need to provide the algorithm
with an initial guess on the size $m_i$ of the blocks. Based on physical
considerations on the expected repartition of the electrons in the
molecule and on the expected localization of the orbitals, the sizes
were fixed  
to values indicated in Table~\ref{table}. The specific block $C_i$ is
then initialized in one of the following three  manners:
\begin{itemize}
\item strategy ${\mathcal I}_1$: the entries of $C$
  are generated randomly, which of course generically yields a bad
  initial guess way;
\item strategy ${\mathcal I}_2$: each block $C_i$ consists of the
  lowest $m_i$ (generalized) eigenvectors associated to the corresponding block matrices $H_i$ and
  $S_i$ in  the matrices $H$ and $S$, respectively. This provides with
  an initial guess, depending on the matrices $H$ and $S$, thus of
  better quality than the random one provided by strategy ${\mathcal I}_1$;  
\item strategy ${\mathcal I}_3$:  the initial guess provided by
  ${\mathcal I}_2$ is optimized with the local fine solver described in
  section~\ref{sec:gendesc}. 
\end{itemize}

\begin{table} \label{table}
\begin{center}
\begin{tabular}{|c|c|c|c|} 
\hline
                & $\mathcal{P}_1$  & $\mathcal{P}_2$   & $\mathcal{P}_3$ \\
\hline
n               &          130     &           200     &  308   \\
q               &    50            &              80   &     126 \\
Bandwith of $S$   &  59      & 79      & 111    \\
Bandwith of $H$   &   99     &   159    &   255  \\
Cut-off for entries of $H$ &    $10^{-12}$ &   $10^{-12}$  & $10^{-10}$    \\
Cut-off for entries of $D$ &  $10^{-11}$ &   $10^{-11}$  &    $10^{-7}$  \\
Size of first  block& $m_1=67$ &$m_1=105$&$m_1=136$\\
Size of  last block& $m_p=67$ &$m_p=106$& $m_p=137$\\
Size of a generic block & $m_i=56$&$m_i=84$&$m_i=104$\\
\hline
\end{tabular}
\caption{Localization parameters and initial size of the blocks
                used in the tests}
\end{center}
\end{table}

\paragraph{Implementation details} Exact  diagonalizations in the local
steps are performed with the routine {\it dsbgv.f} from
the  LAPACK
package~\cite{LAPACK}. In the global
step, the resolution 
of the linear system involving the Hessian matrix is
performed iteratively, using  SYMMLQ \cite{paige}.
Diagonal preconditionning
is used to speed up the resolution. 

The calculations have been performed using only  one processor of a
bi-processor Intel Pentium~IV-2.8~GHz.

\paragraph{Criteria for comparison of results}
For assesment of the quality of the results, we have used two criteria,
regarding the ground-state energy  and the ground-state density
matrix, respectively. For either quantity, the reference calculation is
the calculation using the Gaussian package~\cite{GAUSSIAN98}. 
 The quality of the energy is measured
using the relative error  $\dps e_E= \frac{|E -
    E_0|}{|E_0|}$. For evaluation of the quality of the density matrix,
  we use the
  $L^\infty$ matrix norm  
  \begin{equation}
    \label{norme-matrix}
    e_\infty = \sup_{(i,j)\,\hbox{\small \rm s.t.}\,|H_{ij}|
    \leq \varepsilon } \left| D_{ij}-\left[D_\star\right]_{ij}
    \right|, 
  \end{equation}
where we fix $\varepsilon=10^{-10}$. 
The introduction of the norm (\ref{norme-matrix}) is consistent with  the
cut-off performed on the entries of $H$ (thus the exact value of
$\varepsilon$ chosen).
  Indeed, in practice, the matrix $D$ is only used for the calculations of
  various observables (for instance electronic energy and Hellman-Feynman
  forces), 
  all of the form  $\tr(AD)$ where the symmetric matrix
  $A$ shares  the same pattern as the matrix $H$ (see
  \cite{cances-handbook} for details).  The result is therefore not
  sensitive to entries with indices $(i,j)$ such that
  $|H_{ij}|$ is below the cut-off value.

\subsection{Illustration of the role of the local and global steps}

Our MDD method consists in three
ingredients:
\begin{itemize}
\item the local optimization of each block performed in the local step;
\item the transfer of vectors from some
blocks to other blocks, along with the modification of the block sizes
$m_i$, again in the local step;
\item the optimization performed in the global step.
\end{itemize}
To highlight the necessity of each of the
ingredients, and their impacts on the final result, we compare our
MDD algorithm with 
three simplified variants. Let us denote by
\begin{itemize}
\item strategy ${\mathcal S}_1$: local optimization of the blocks, without
  allowing variations of the block sizes, and no global step;
\item strategy ${\mathcal S}_2$: full local step  (as defined in
  Section~\ref{sec:gendesc}), no global step;
\item strategy ${\mathcal S}_3$: local optimization of the blocks, without
  allowing for variations of the block sizes, and global step;
\item strategy ${\mathcal S}_4$: full algorithm.
\end{itemize}

\medskip

\begin{figure}[h]
\centering
\psfig{figure=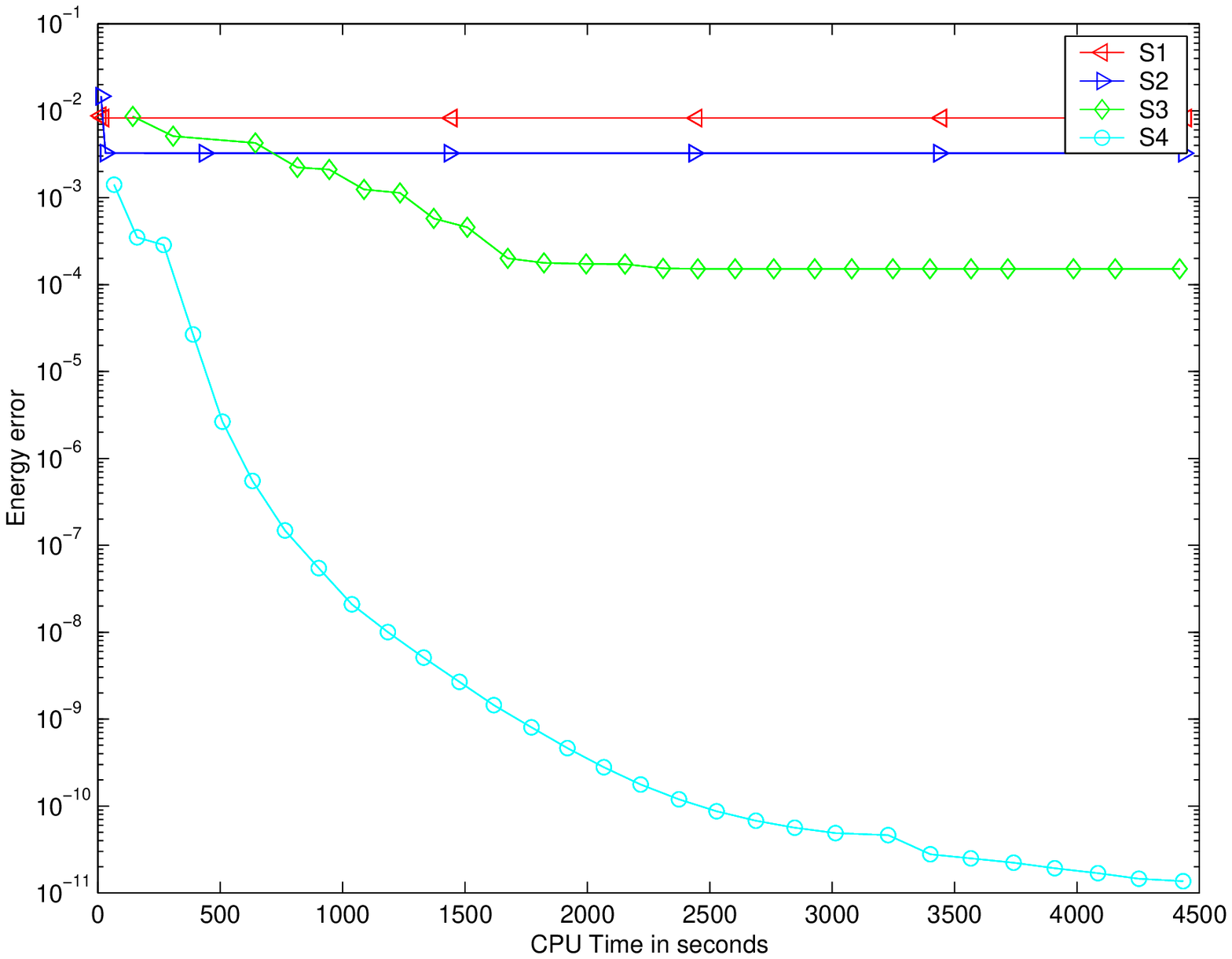,height=8truecm}
\caption{Energy error versus CPU time obtained
  with a bad initial guess (${\mathcal I}_1$).}
\label{fig:res1}
\end{figure}

\begin{figure}[h]
\centering
\psfig{figure=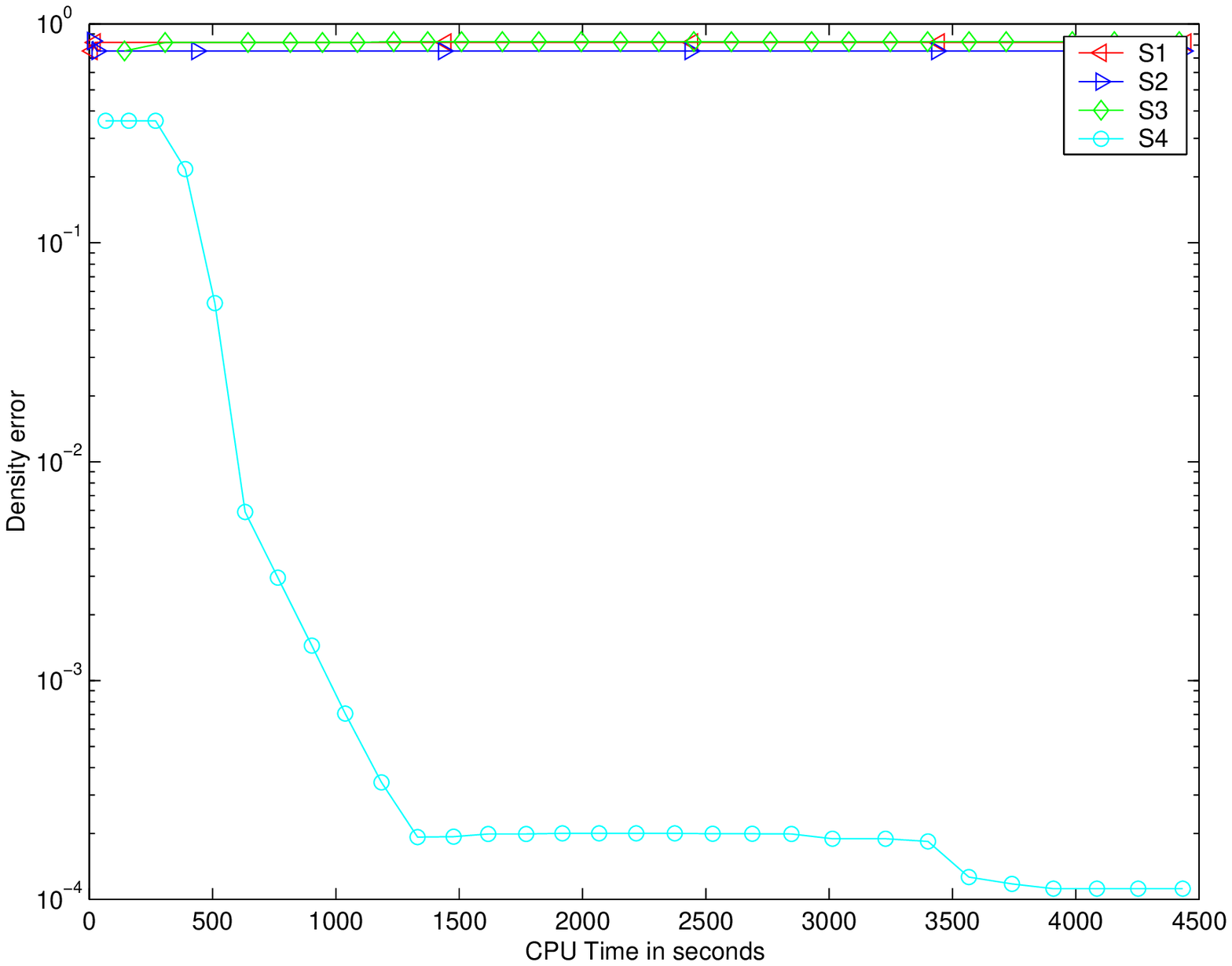,height=8truecm}
\caption{Density error versus CPU time obtained
  with a bad initial guess (${\mathcal I}_1$).} 
\label{fig:res2}
\end{figure}

\begin{figure}[h]
\centering
\psfig{figure=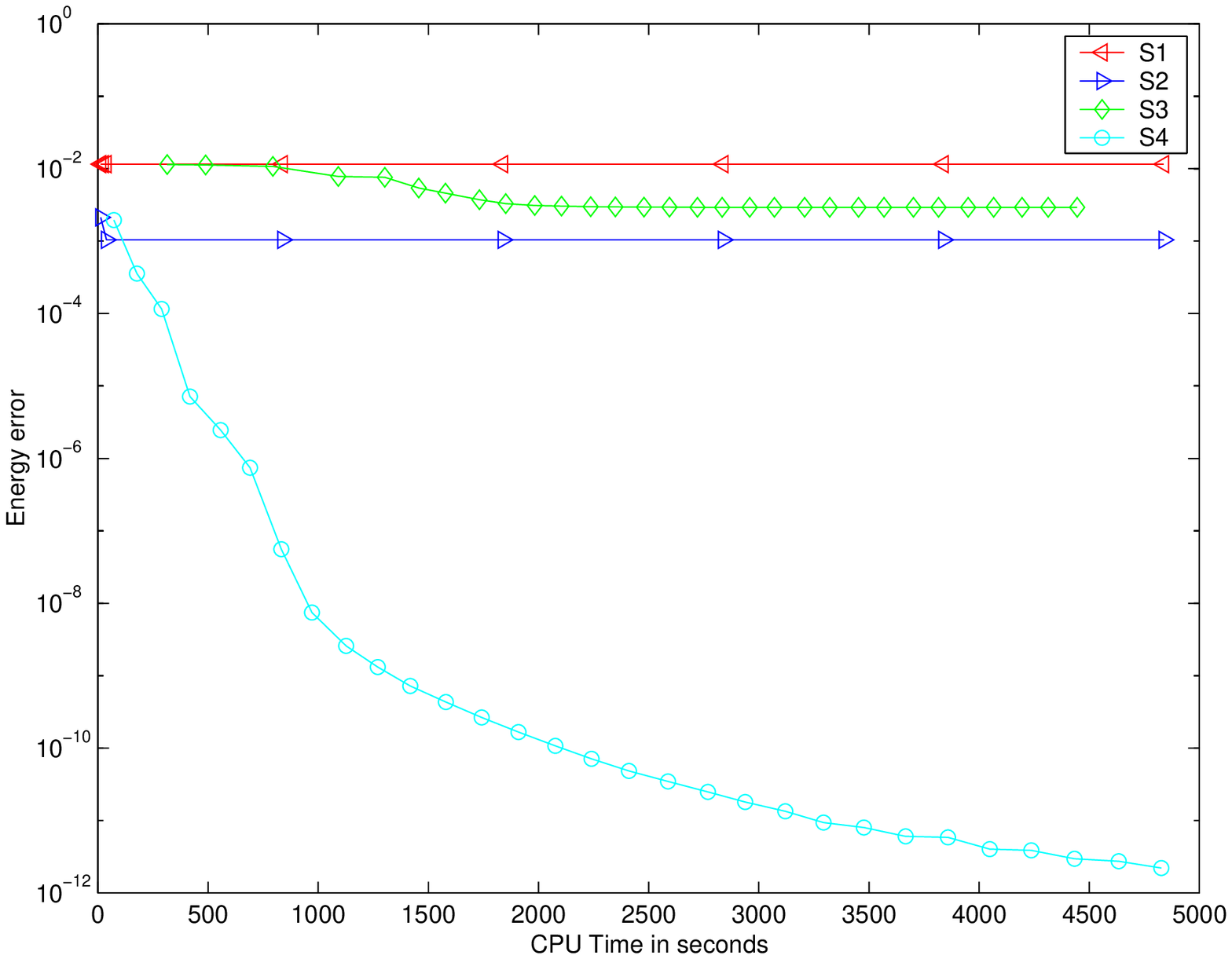,height=8truecm}
\caption{Energy error versus CPU time obtained
  with a better initial guess (${\mathcal I}_2$).} 
\label{fig:res3}
\end{figure}

\begin{figure}[h]
\centering
\psfig{figure=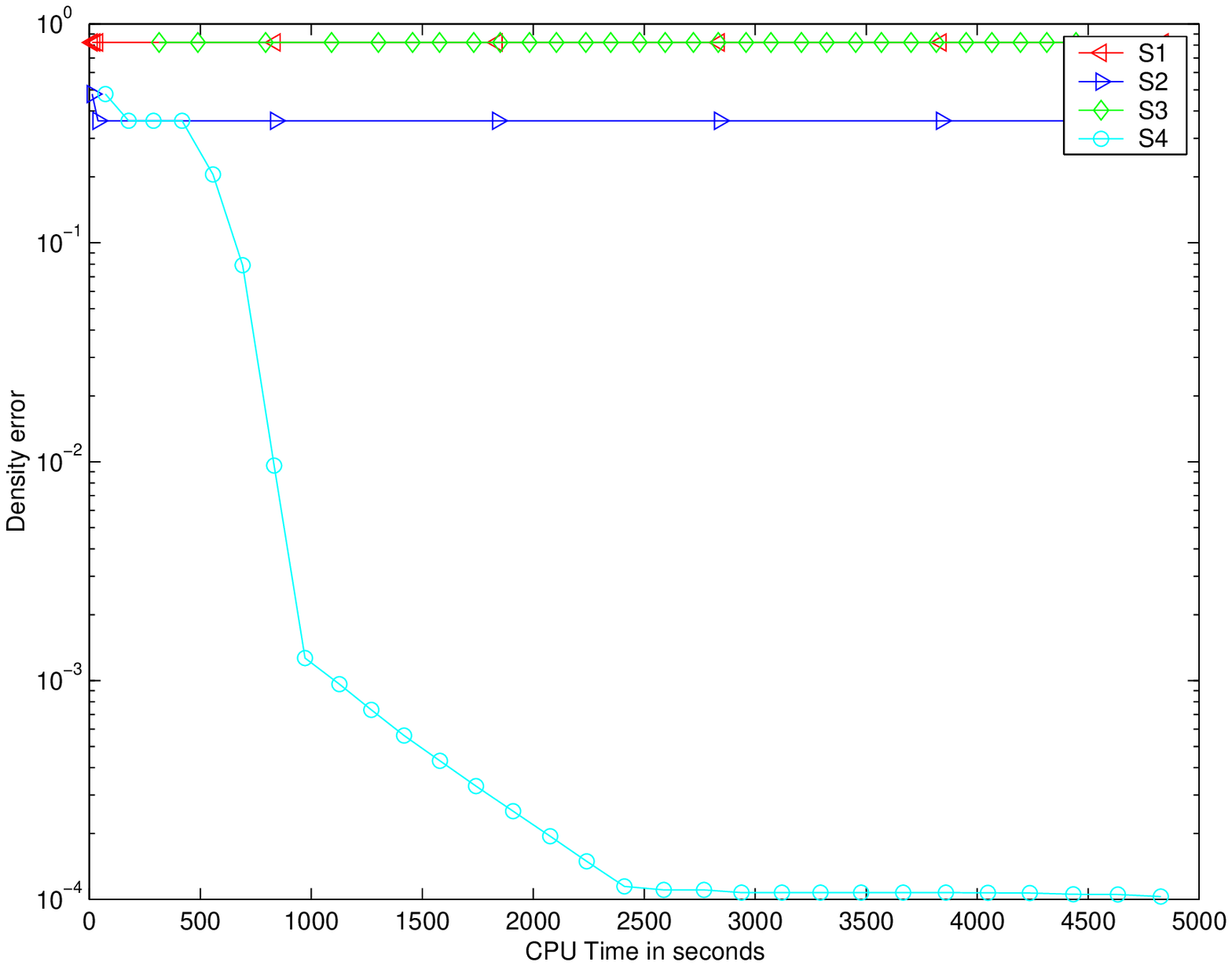,height=8truecm}
\caption{Density error versus CPU time obtained
  with a better initial guess (${\mathcal I}_2$).}
\label{fig:res4}
\end{figure}

We compare the rate of convergence for the above four strategies. Two
categories of tests are performed, depending on the quality of the
initial guess. The results displayed on Figures~\ref{fig:res1} to
~\ref{fig:res4} concern polymer  $\mathcal{P}_1$ with $n_m=801$
monomers. This corresponds to 
$N_b = 8050$ and $N=5622$. Analogous tests were performed on
$\mathcal{P}_2$  and $\mathcal{P}_3$, but we do not present them here,
for brevity.

 The energy of the ground
state of this matrix (i.e. the minimum of (\ref{eq:infF})) is $E_0 =
-27663.484$. The number of blocks considered is $p=100$. For the
global step, we have collected these 100  blocks in $99$ overlapping
groups of $2$ blocks. Interestingly,  such a
partition provides with optimal results regarding CPU time and memory requirement.
It is observed on Fig.~\ref{fig:res1}-\ref{fig:res4} that ${\mathcal
  S}_1$, ${\mathcal S}_2$ and ${\mathcal S}_3$ are not 
satisfactory for they converge towards some local, non global, minima of
(\ref{eq:infF-blocks}) whatever the initial guess. The failure of the
strategy $S_3$  performed on the initial guess ${\mathcal I}_2$ is
surprising: this initial guess is not good enough. Indeed, if the
initial guess is  ${\mathcal I}_3$, we check numerically that the strategies 
${\mathcal S}_3$ and ${\mathcal S}_4$ behave identically. Notice that
the strategy ${\mathcal I}_3$ is identical to ${\mathcal S}_2$ applied to the initial guess  ${\mathcal I}_2$.

We also remark  that the strategy ${\mathcal S}_4$ performs very well
whatever the initial guess (see Fig.~\ref{fig:res2} and
Fig.~\ref{fig:res4}). The same behavior is observed for the polymers
${\mathcal P}_2$ and ${\mathcal P}_3$. 
Finally, after orthonormalization, the Density Matrix Minimization (DMM)
method~\cite{DMM} 
failed with the random initial guess and reveals very slow with the
initial guess  ${\mathcal I}_2$. That is the reason  why we consider the initial guess  ${\mathcal I}_3$ to compare these methods.

\medskip

\subsection{Comparison with two other
  methods}

Having emphasized the usefulness of all the ingredients of our
MDD algorithm, we now compare it to two other algorithms:
\begin{itemize}
\item the diagonalization routine {\it dsbgv.f} from the LAPACK library;
\item the Density Matrix Minimization (DMM) method~\cite{DMM}.
\end{itemize}
These two algorithms are seen as prototypical approaches for standard
diagonalization algorithms and linear scaling techniques respectively.
They are only  used here for  comparison purposes.
Regarding linear
  scaling methods, two other popular approaches are  the Fermi Operator
  method \cite{FOE} and  the McWeeny
  iteration method  \cite{McWeeny}. We have observed that, at least in
  our own implementation, based on the literature, they are
  outperformed by the DMM method for the actual chemical systems we have
  considered. We therefore take DMM as a reference method for our
  comparison. 

\medskip

Recall that the routine {\it dsbgv.f} consists in the three-step procedure
\begin{itemize}
\item transform the generalized eigenvalue problem into a standard
  eigenvalue problem by applying a Cholesky factorization to $S$;
\item reduce the new matrix to be diagonalized to a tridiagonal form;
\item compute its eigenelements by using the
  implicit $QR$ method.
\end{itemize}
The algorithmic complexity of this approach is in $N_b^3$ and the
required memory scales as $N_b^2$. 

For the description of DMM method, we refer to \cite{DMM}. Let us only
mention here that this approach consists in a minimization procedure,
applied to the energy expressed in terms of the density matrix. 
Both the   algorithmic complexity and the memory needed for performing the DMM approach scale linearly with
respect to the size  $N_b$ of the matrix.  
The DMM method is initialized with the density matrix $D=CC^t$
computed with  the initial guess $C$ of the domain decomposition method. 
Two important points for the tests shown below   are the
following. 

\begin{figure}[h]
  \centering
   \epsfig{figure=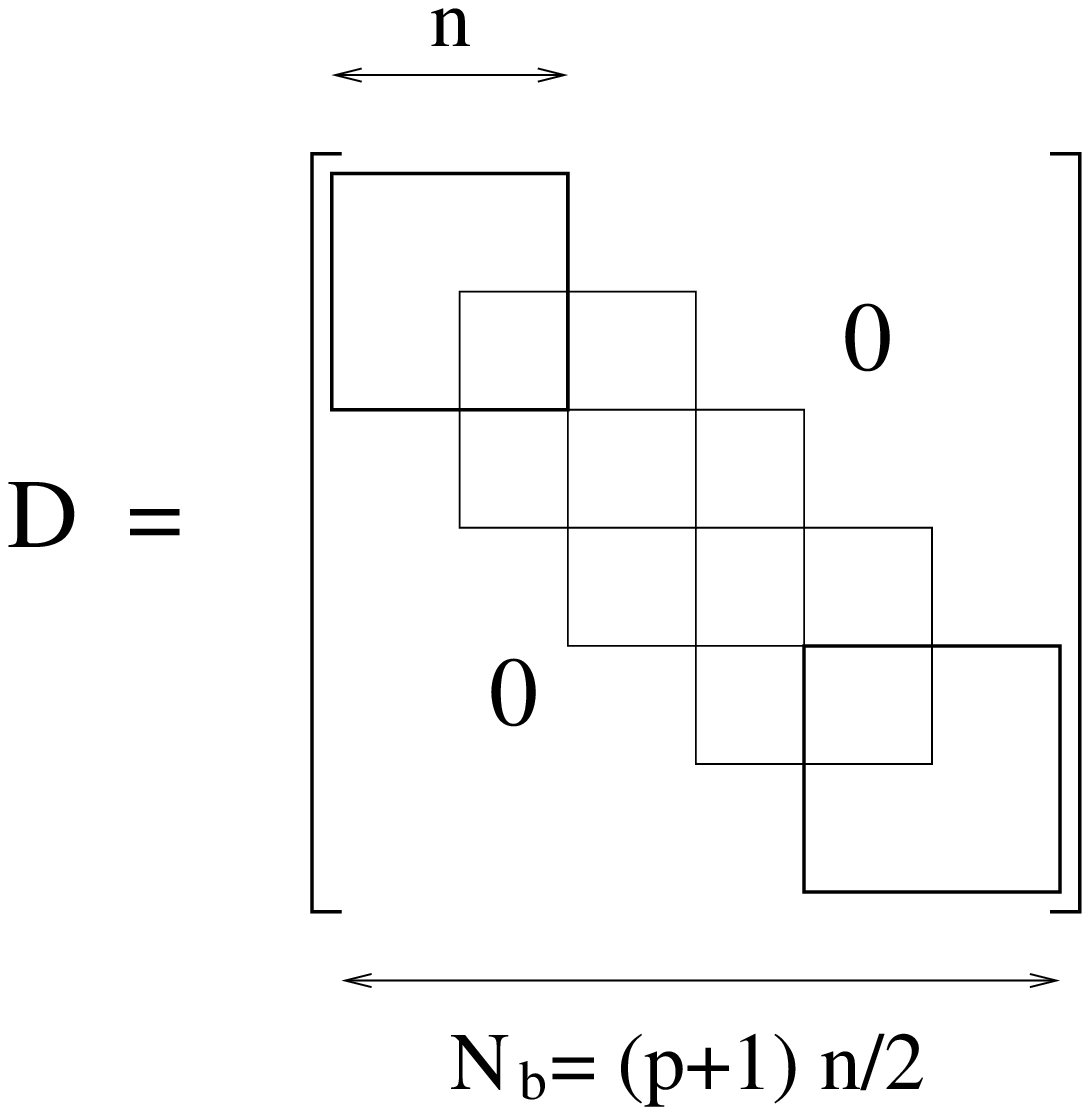,height=5truecm}
    \caption{
     Block structure of the matrix $D$.\label{fig:D}}
\end{figure}

First,  we perform a cut-off on the coefficients
on the various matrices manipulated throughout the calculation: only the
terms of the density matrices within  
the frame defined in Figure~\ref{fig:D} are taken into account. Such a
cut-off has some impact 
on the qualities of the results obtained with the DMM method. We are
however not able to design a better comparison. 

Second,  the DMM method requires the knowledge of
  the Fermi level (as is the case for the  linear scaling methods
  commonly used in practice to date). The determination of the Fermi
  level is the purpose of an outer optimization loop. In contrast, the
  MDD approach computes an approximation of the Fermi
  level at each iteration. Here, for the purpose
of comparison, we \emph{provide} DMM with the exact value of the Fermi
level.  Consequently, the CPU times for the DMM method displayed in the
  sequel are underestimated. 

\medskip

We emphasize that the routine {\it dsbgv.f} computes the entire
spectrum of the matrix, both eigenvalues and eigenvectors. In contrast, the MDD
approach only provides with the lowest $N$ eigenvalues, 
among $N_b$, and
the projector on the vector space spanned by the
corresponding eigenvectors, not the eigenvectors themselves.

\subsubsection{Comparison with Direct diagonalization and DMM}

We have computed the ground states of the polymers ${\mathcal P}_1$,
${\mathcal P}_2$ and ${\mathcal P}_3$ with the three methods (direct
diagonalization, DMM and MDD) and for various numbers $n_m$ of
monomers, corresponding to matrix sizes $N_b$ in the range $10^3$-$10^5$.

\begin{figure}[ht]
\centering
\psfig{figure=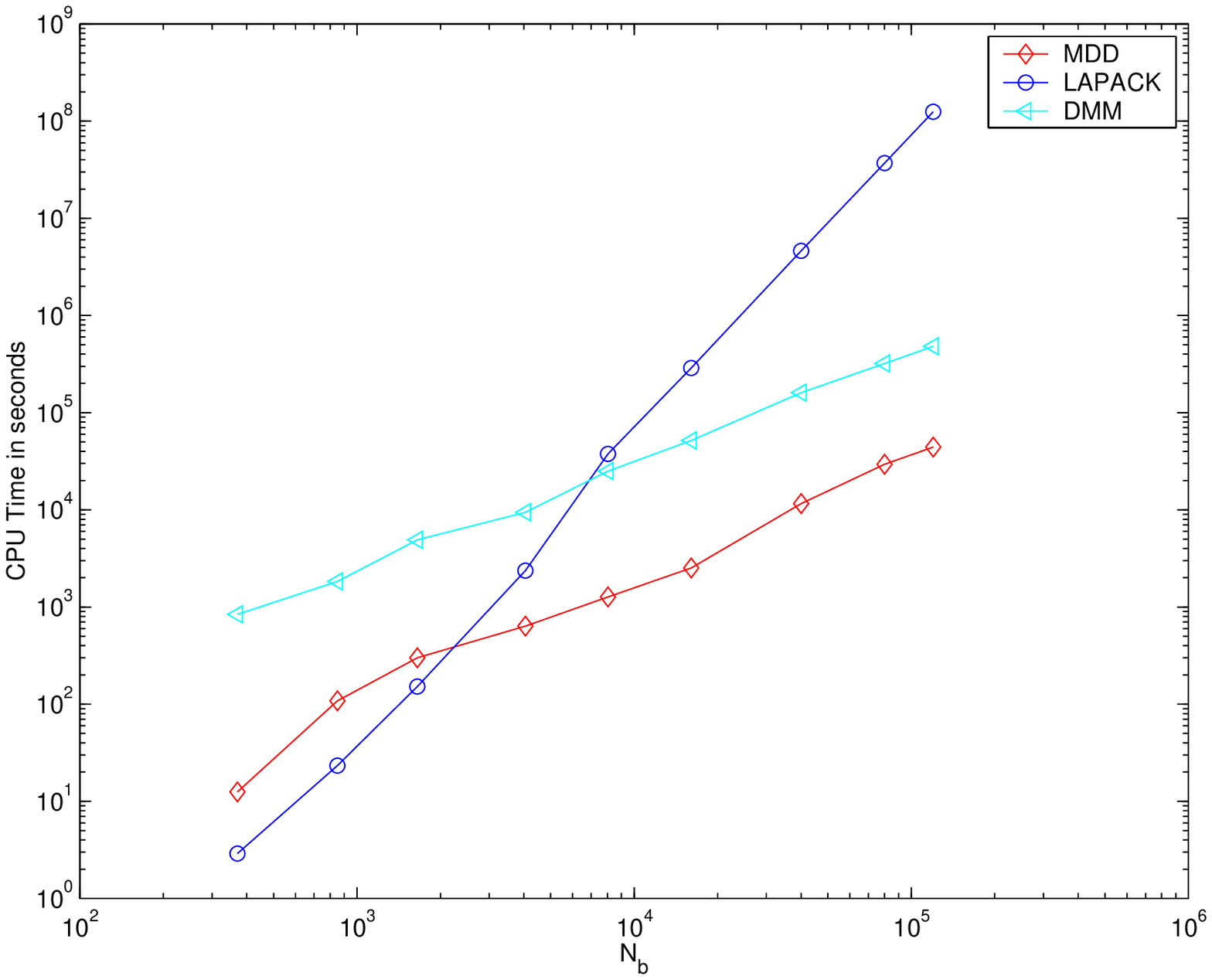,height=8truecm}
\psfig{figure=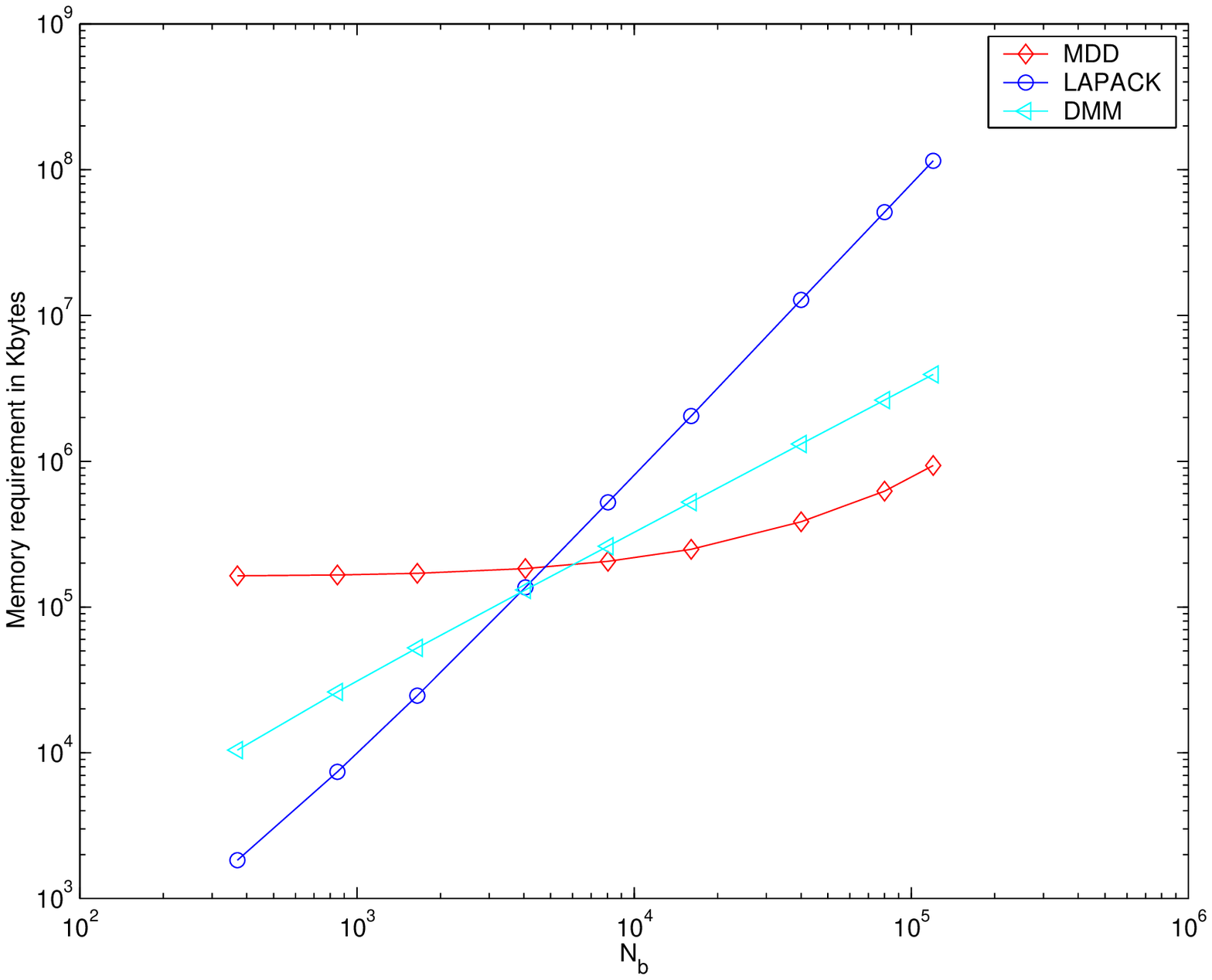,height=8truecm}
\caption{Scaling of the CPU time (top) and memory requirement (bottom)
  for the polymer ${\mathcal P}_1$.} 
\label{fig:res5}
\end{figure}

\begin{figure}[h]
\centering
\psfig{figure=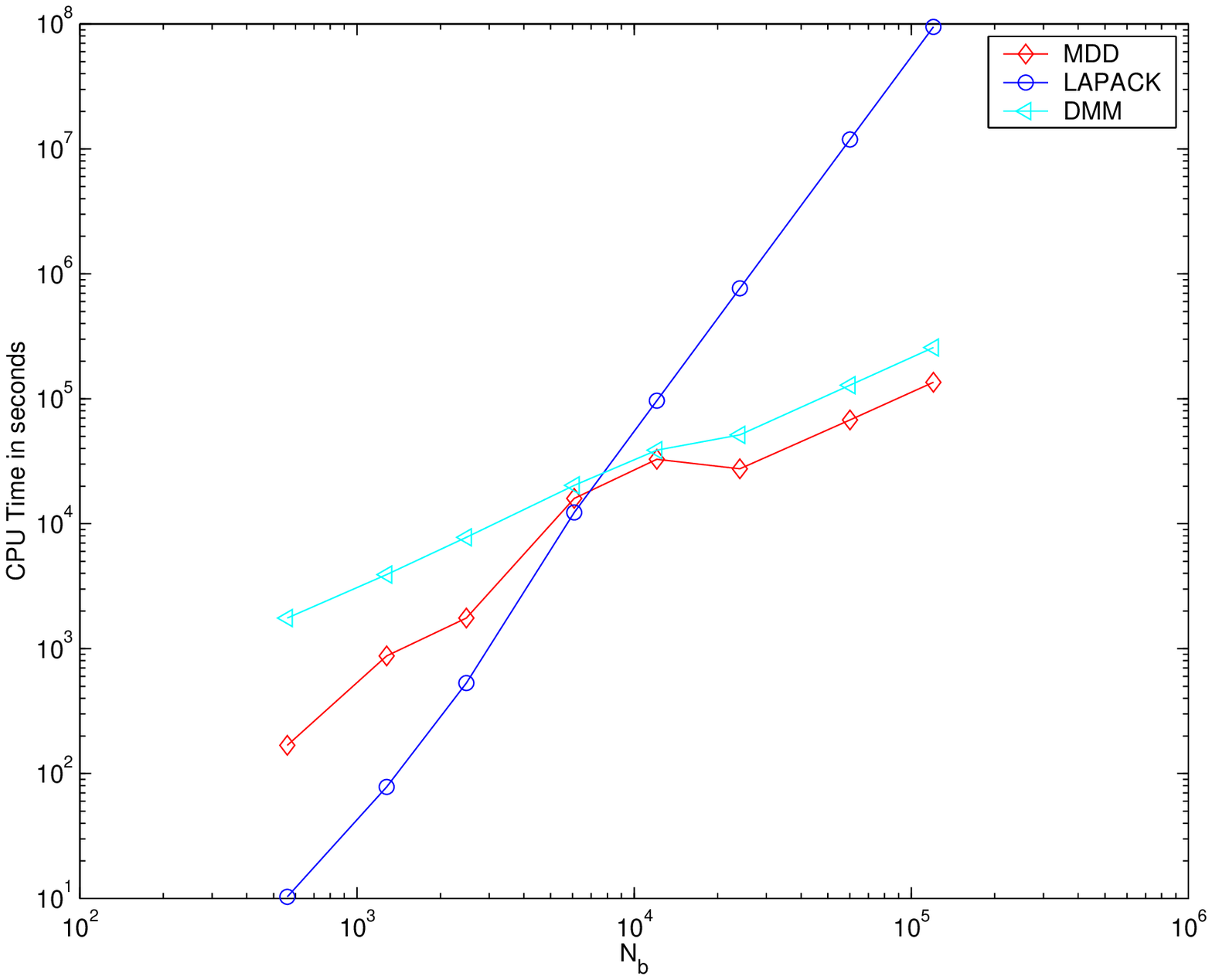,height=8truecm}
\psfig{figure=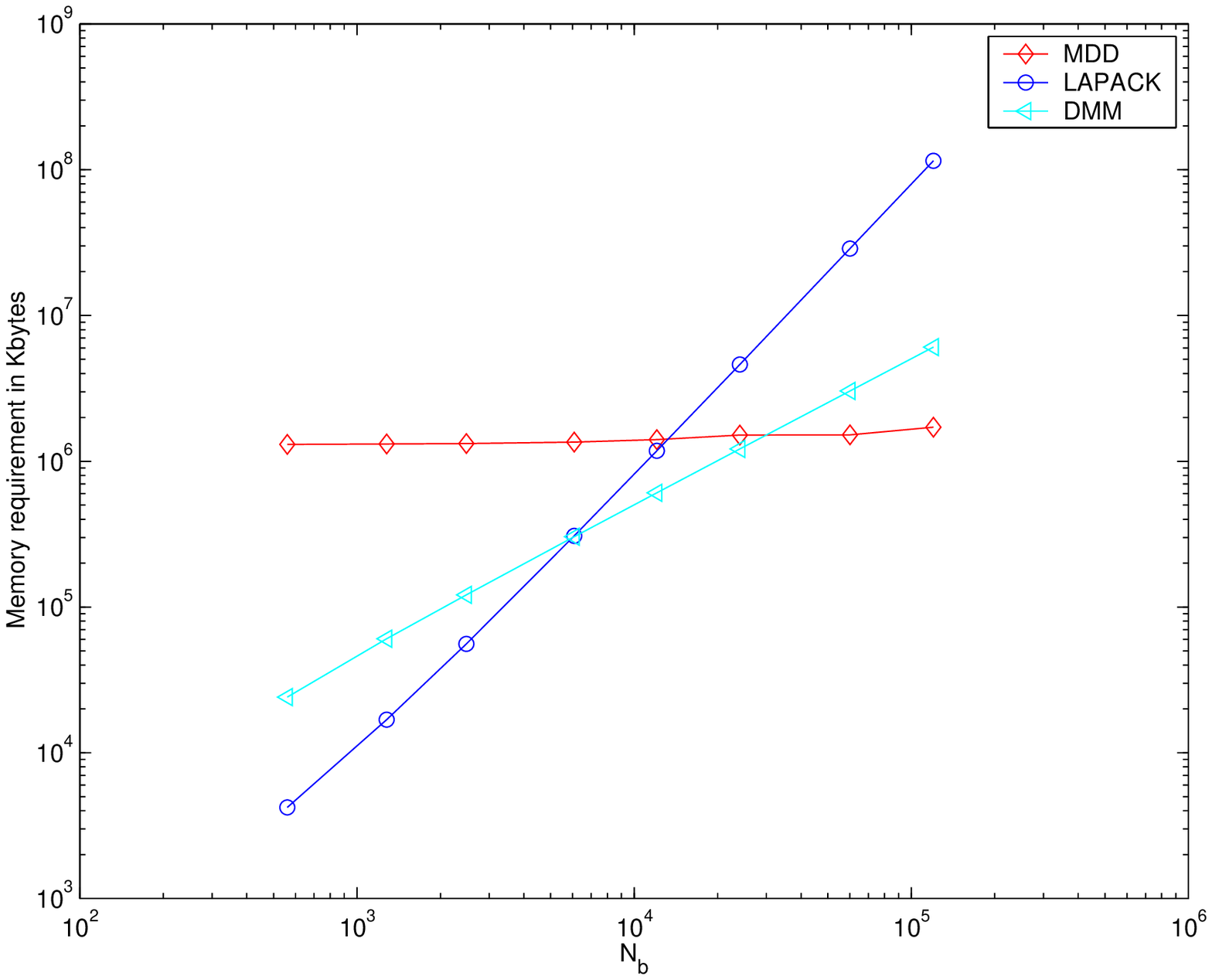,height=8truecm}
\caption{Scaling of the CPU time (top) and memory requirement (bottom)
  for the polymer ${\mathcal P}_2$.}
\label{fig:res7}
\end{figure}

\begin{figure}[h]
\centering
\psfig{figure=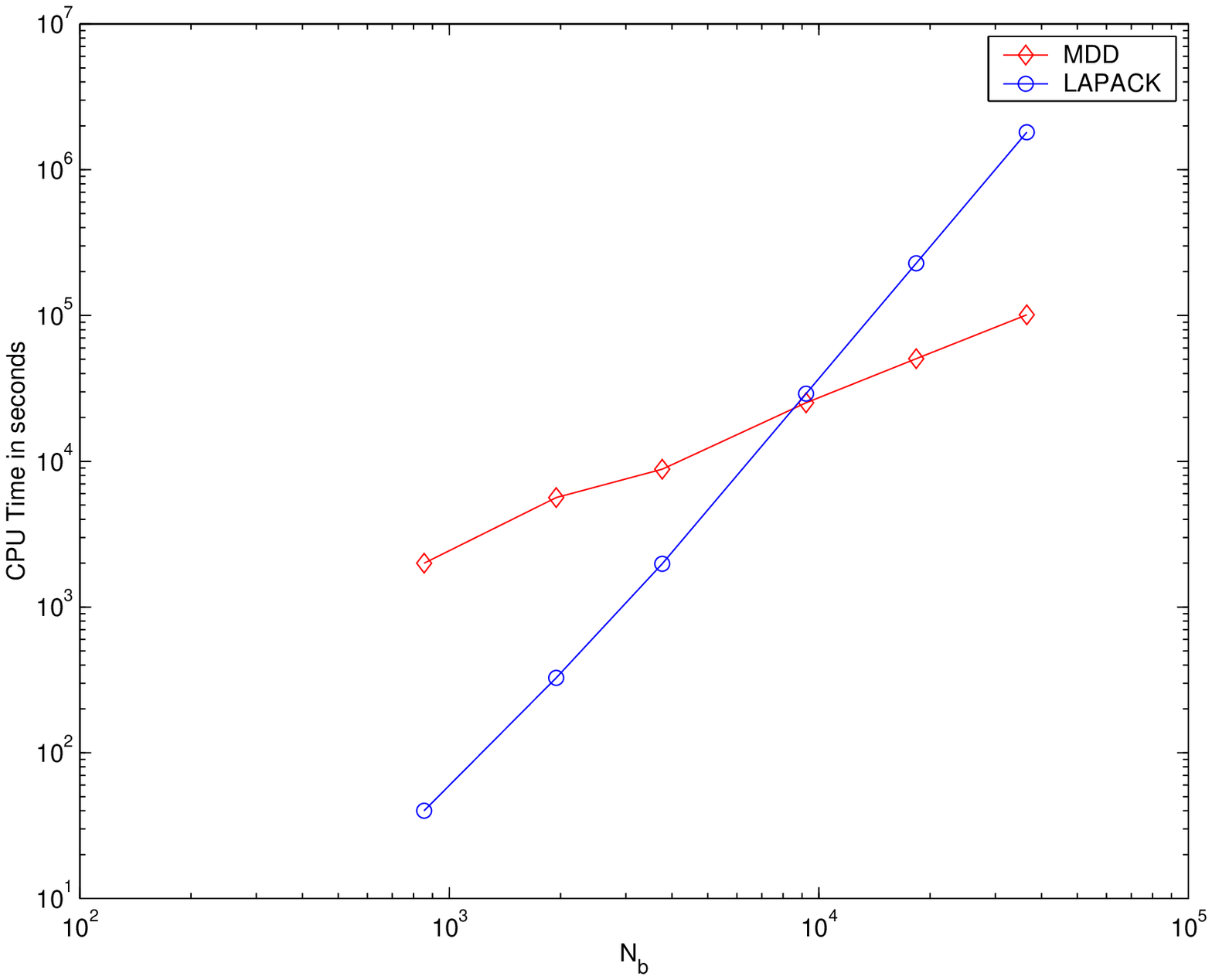,height=8truecm}
\psfig{figure=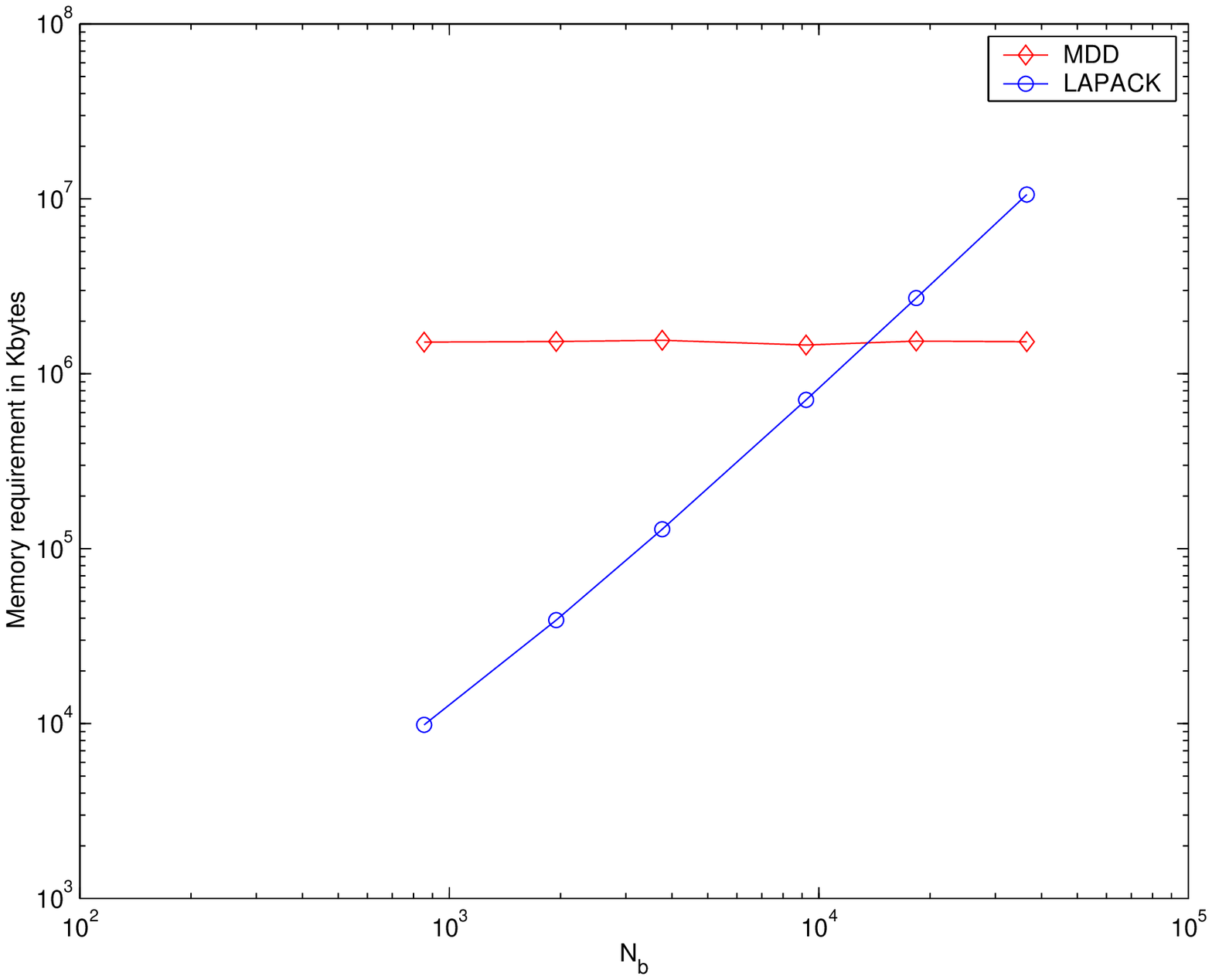,height=8truecm}
\caption{Scaling of the CPU time (top) and memory requirement (bottom)
  for the polymer ${\mathcal P}_3$.}
\label{fig:res9}
\end{figure}

For DMM and MDD,  the initial guess is generated following the
strategy ${\mathcal I}_3$. The results regarding the CPU time at convergence and the memory
requirement are displayed on
Figures~\ref{fig:res5} to~\ref{fig:res9} for the polymers
$\mathcal P_1$, $\mathcal P_2$, and $\mathcal P_3$ respectively.

For small values of $N_b$, i.e. up to around $10^4$, the results
observed for the direct diagonalization, DMM and MDD agree.  The
CPU times for our MDD approach scale linearly with $N_b$.

For larger values of $N_b$, the limited memory prevented us from either
performing an exact diagonalization or from implementing DMM.
So, we extrapolate the CPU time and
memory requirement according to the scaling observed for smaller
$N_b$.

The data for the DMM method are not plotted in
Figure~\ref{fig:res9} as the DMM method does not
converge for the polymer ${\mathcal P}_3$ when the number of monomers 
exceeds $10^3$. From our point of view, it comes from the truncation
errors which cause the divergence of the method (note that the truncation
strategy we consider here is very simple).  

\subsubsection{Comparison with DMM and a hybrid strategy}

We now concentrate on the two approaches that scale linearly, namely DMM
and MDD. 
We consider 
\begin{itemize}
\item ${\mathcal P}_1$ with $4001$ monomers, corresponding to $N_b=40050$,
\item ${\mathcal P}_2$ with $2404$ monomers, corresponding to $N_b=24080$,
\item ${\mathcal P}_3$ with $208$ monomers, corresponding to $N_b=854$.
\end{itemize}
These particular values have been chosen for the purpose of having
simple values for the numbers of blocks.
For each of the three polymers, we compare the DMM and MDD methods
initialized by 
the strategy ${\mathcal I}_3$ and a hybrid strategy. The hybrid strategy
consists of a certain number of iterations performed with  MDD, until
convergence is reached for this method, followed
by iterations with 
DMM. We use the following stopping criterion for MDD:
\begin{equation} \label{eq:stopping}
\dps \|D_n-D_{n-1}\| \geq \|D_{n-1}-D_{n-2}\| \quad
\mbox{\textrm{and}} \quad \|D_n-D_{n-1}\| \leq \epsilon_a 
\end{equation} 
where $\epsilon_a$ is a threshold parameter. We take
$\epsilon_a=10^{-4}$, respectively $\epsilon_a=10^{-3}$, for the polymer
${\mathcal P}_1$, respectively ${\mathcal P}_2$ and ${\mathcal P}_3$. 

\begin{figure}[h]
\centering
\psfig{figure=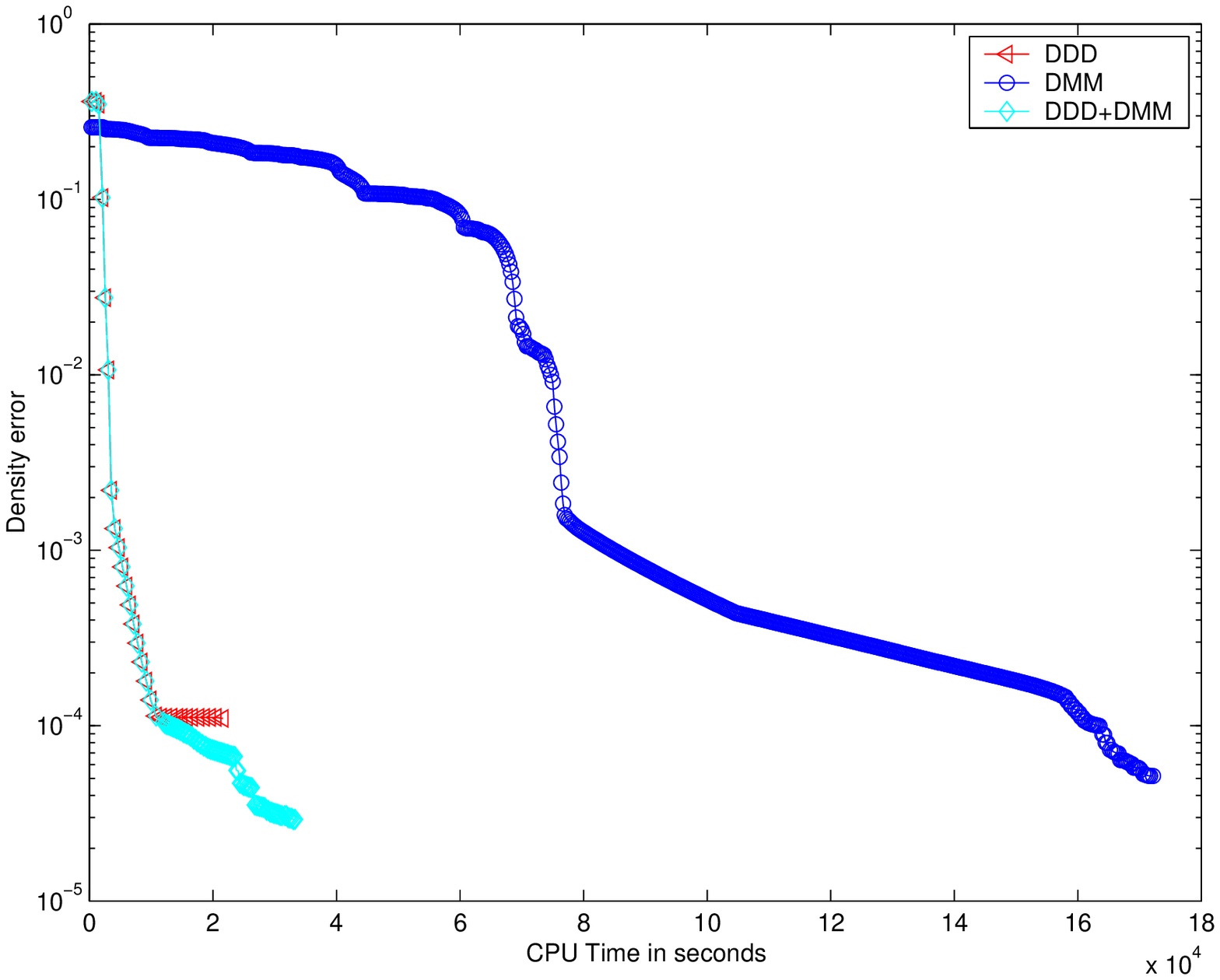,height=8truecm}
\caption{Evolution of the density error with the CPU time for the polymer ${\mathcal P}_1$ made of $4001$ monomers.}
\label{fig:res11}
\end{figure}

\begin{figure}[h]
\centering
\psfig{figure=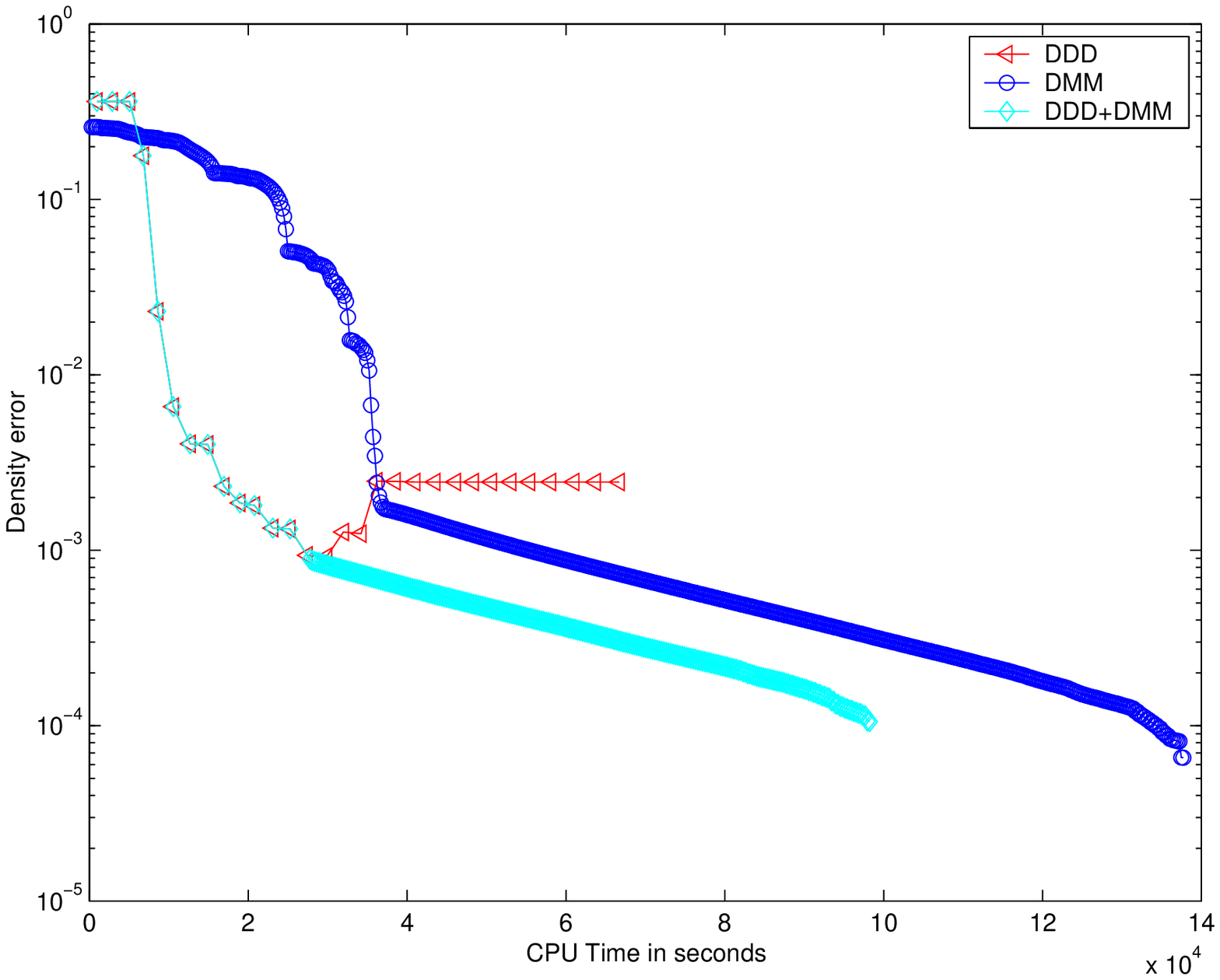,height=8truecm}
\caption{Evolution of the density error with the CPU time for the polymer ${\mathcal P}_2$ made of $2404$ monomers.}
\label{fig:res12}
\end{figure}

\begin{figure}[h]
\centering
\psfig{figure=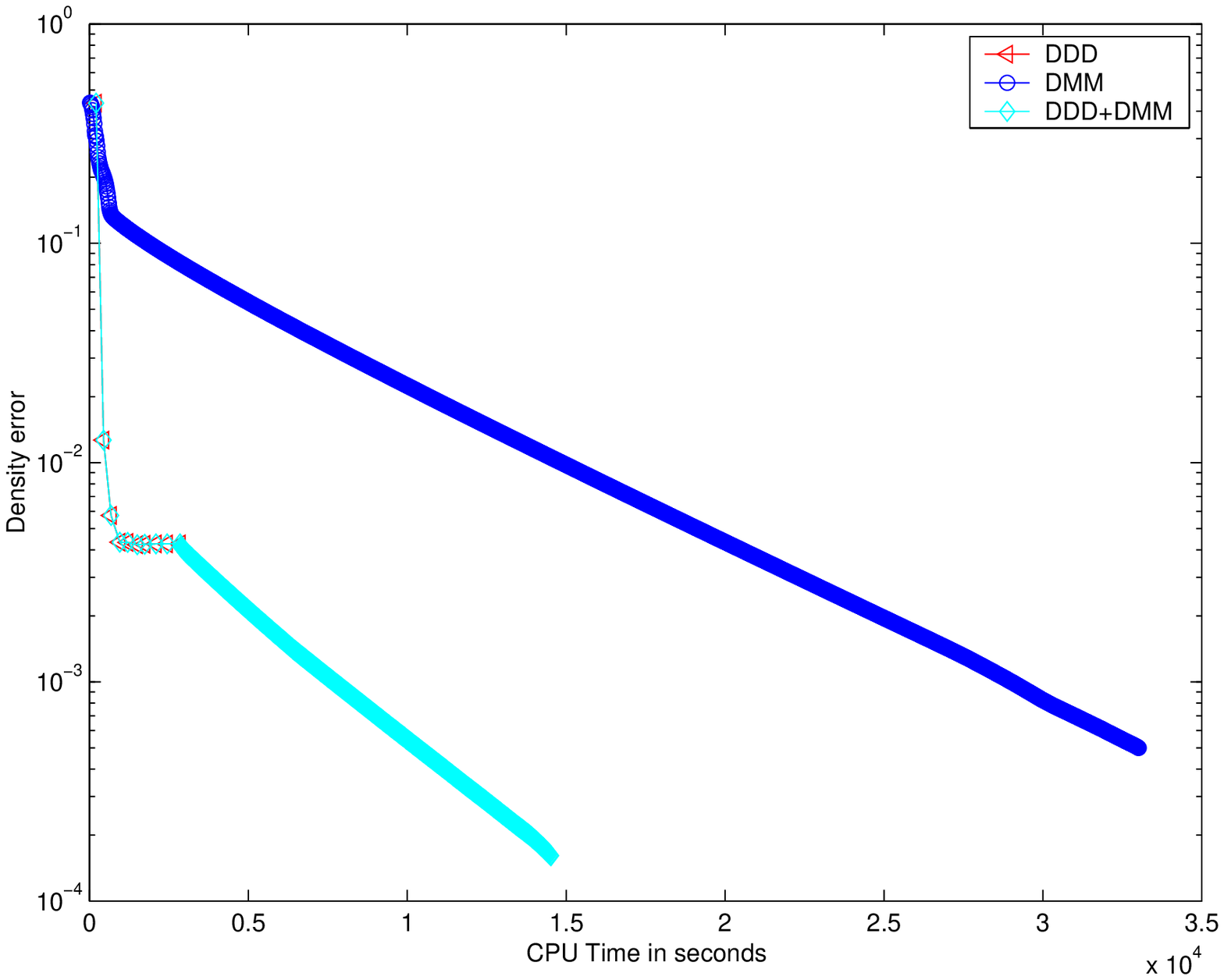,height=8truecm}
\caption{Evolution of the density error with the CPU time for the
  polymer ${\mathcal P}_3$ made of $208$ monomers.} 
\label{fig:res13}
\end{figure}

The Figures~\ref{fig:res11} to~\ref{fig:res13} show the evolution of the
error in density versus  CPU time.
The hybrid version is demonstrated to be a  very efficient combination
of the two algorithms.

For completeness, let us highlight the temporary  increase for the error in
density appearing in Fig.~\ref{fig:res12} when MDD is used on $\mathcal
P_2$. Analogously, the energy of the current solution, which is actually
below the reference energy,  also increases. In
fact, this is due to a loss of precision in the orthonormality
constraints. In MDD, these constraints are not imposed exactly at each
iteration, but only approximately (see equation~\ref{eq:epsilonL}). 

\begin{figure}[h]
\centering
\psfig{figure=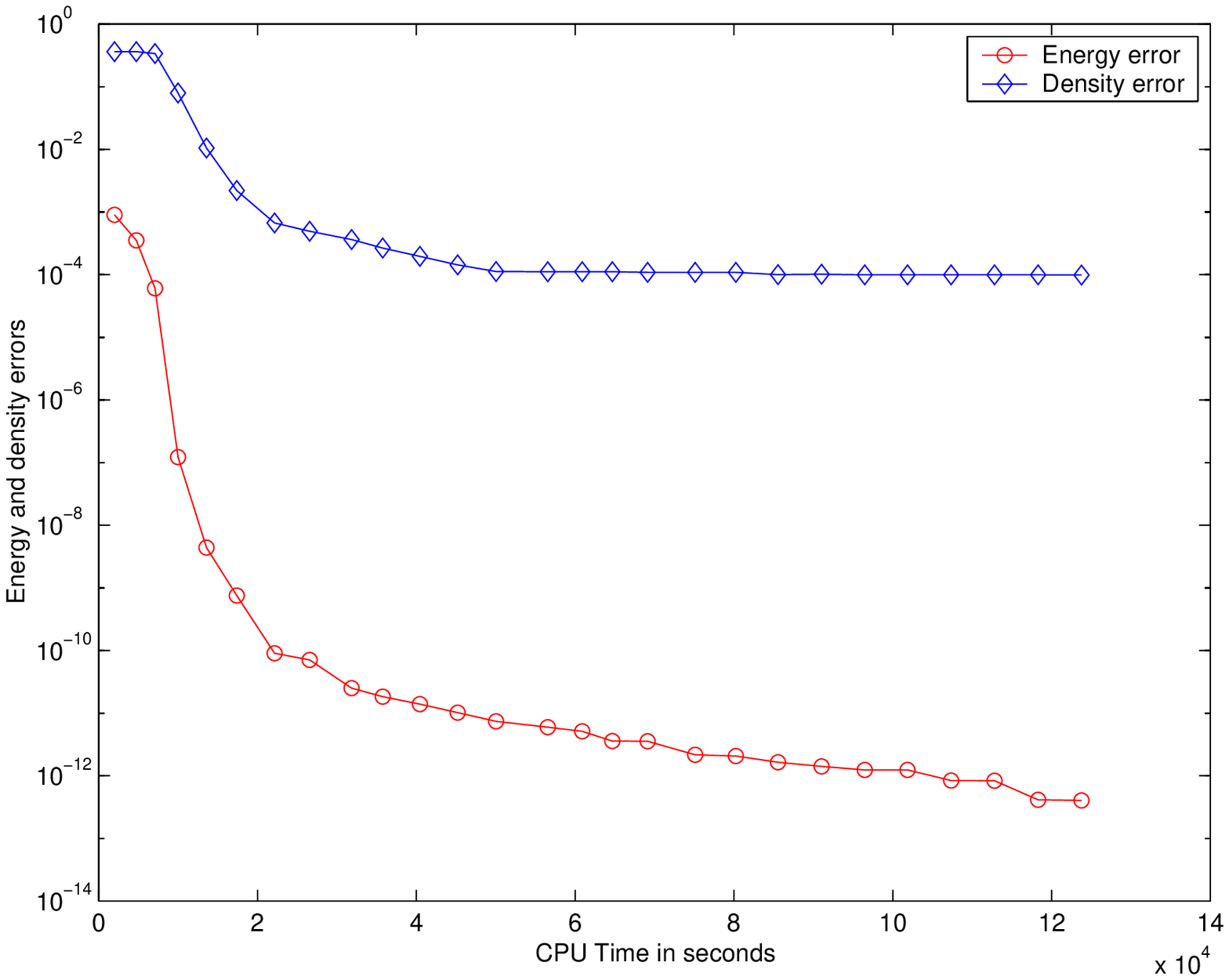,height=8truecm}
\caption{Evolution of the MDD energy and density errors versus CPU time for
  the polymer ${\mathcal P}_1$ ($20001$ monomers, $N_b=200050$).}
\label{fig:res14}
\end{figure}

\begin{figure}[h]
\centering
\psfig{figure=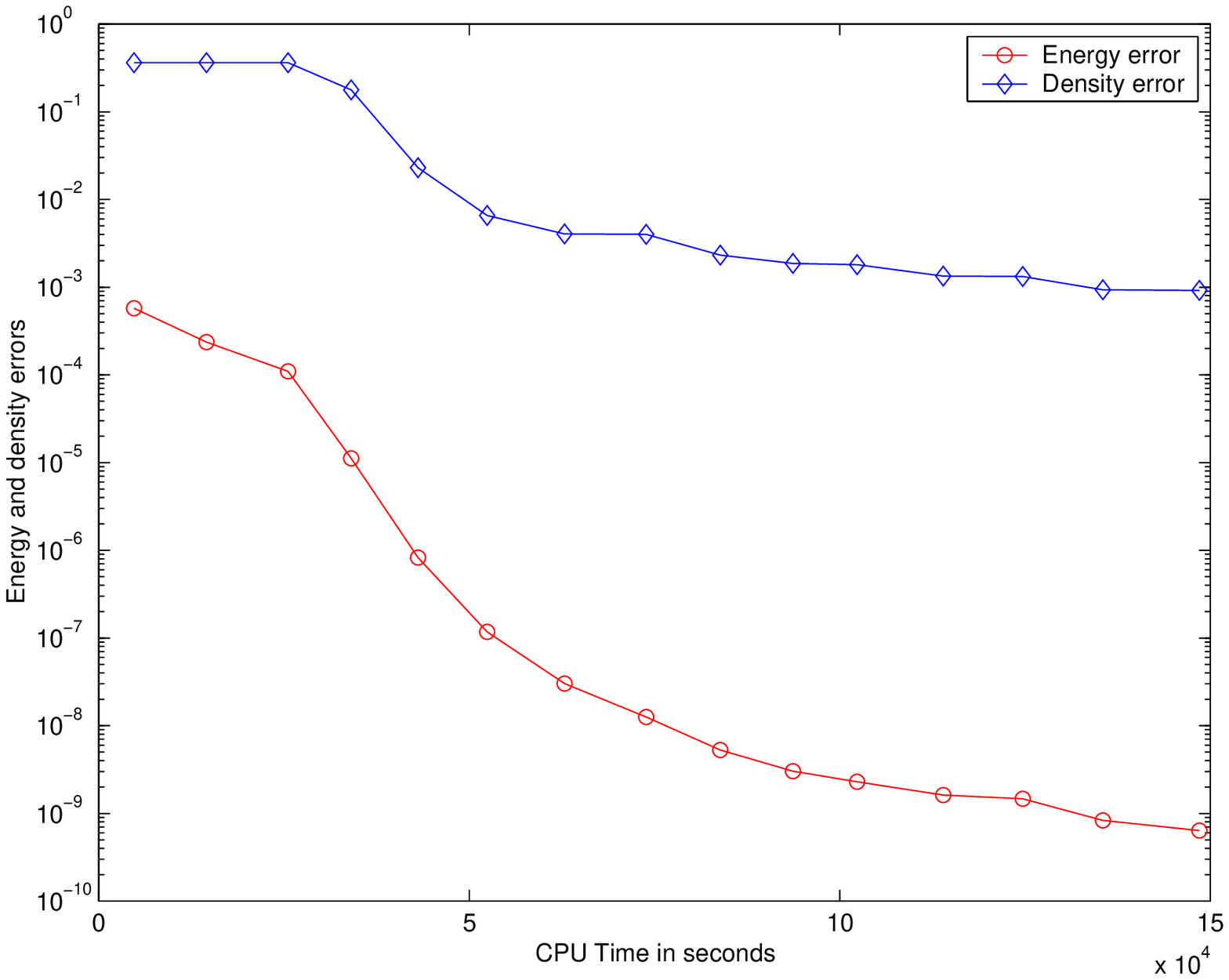,height=8truecm}
\caption{Evolution of the MDD energy and density errors versus CPU time for
  the polymer ${\mathcal P}_2$ ($12004$ monomers, $N_b=120080$).}
\label{fig:res15}
\end{figure}

\begin{figure}[h]
\centering
\psfig{figure=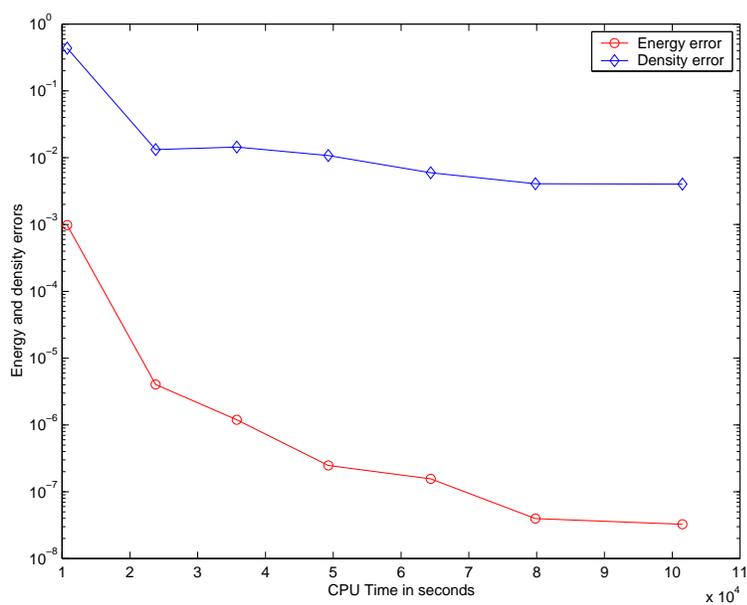,height=8truecm}
 \caption{Evolution of the MDD energy and density errors versus CPU time for
   the polymer ${\mathcal P}_3$ ($5214$ monomers, $N_b=36526$).}
\label{fig:res16}
\end{figure}

Finally, we report in figures~\ref{fig:res14} to~\ref{fig:res16} the
results obtained with MDD for the largest possible case that can be
perfomed on our platform, owing to memory limitation. We used the initial
guesses obtained with the
strategy ${\mathcal I}_3$. Notice that
for the local step the memory requirement scales linearly with respect
to the number $n_m$ of monomers, 
while for the global step, the memory requirement is independent of
$n_m$. Therefore, for 
large polymers,  the memory needed by MDD is controled by the local
step. In contrast, for small polymers, the most demanding step in terms
of memory is the global step. 

\section{Conclusions and remarks}

The domain decomposition algorithm introduced above performs
well, in comparison to the two standard methods considered. More
importantly, our approach is an effective \emph{preconditionning
  technique} for DMM iterations. 
Indeed,  MDD provides a rapid and  accurate approximation, both
in terms of  energy and  density matrix, regardless of the quality
of the initial guess. In contrast,  DMM  outperforms MDD when
the initial guess is good, but only performs poorly, or may  even diverge,  when this is not the
case.  The combination of the two methods seems to be optimal. More generally,
our MDD algorithm could constitute a good preconditionner to all variational
methods, such as  the Orbital Minimization method \cite{Siesta}.

\medskip

 Regarding the comparison with DMM, the following comments are
 in order.

 \begin{itemize}
\item 
  All our calculations have been performed on a single processor
machine. Potentially, both DMM and MDD  should
exhibit the same speed-up when parallelized. We therefore consider
 the comparison valid, at least qualitatively,  for parallel
implementations. The parallelization of the MDD is currently in
progress, and hopefully will confirm the efficiency of the approach.

\item 
We recall the Fermi level has to be provided to the DMM method. This is
an additional argument in favor of the MDD approach.

\item 
 The MDD method, in contrast to  the other linear
 scaling methods, does not perform any truncation in the
 computations. So, once the profile of $C$ is choosen, the method does not
 suffer of any  instabilities, contrary to DMM (or OM) for which
 divergences have been observed for the polymer ${\mathcal P}_3$. 

\item 
 The domain decomposition method makes use of several threshold
 parameters. For the three polymers we have
 considered, the optimal values of these parameters, except for the
 stopping criterion $\epsilon_a$ (equation (\ref{eq:stopping})), are the
 same. We do not know yet if this interesting feature is a general
 rule. 

\item 
Recall our method solves problem~(\ref{eq:infF-blocks}), which
is only an 
approximation of problem~(\ref{eq:infF2}). Therefore, the relative error
obtained in the limit is only a measure of the difference between
(\ref{eq:infF-blocks}) and~(\ref{eq:infF2}). In principle, such
a difference could be made arbitrarily small by an appropriate choice of
the parameters of problem~(\ref{eq:infF-blocks}).

\item Finally, let us emphasize that there is much room for improvement in
both the local and the global steps. We have designed an overall multilevel
strategy that performs well, but each subroutine may be significantly
improved. Another interesting issue is the interplay between the
nonlinear loop in the Hartree-Fock or Kohn-Sham problems
(Self-Consistent Field - SCF - convergence
\cite{cances-handbook,outperform,EDIIS}) and the linear subproblem
considered in the present article. Future efforts will go in these
directions. 
 \end{itemize}

\bigskip

\noindent
{\sc Acknowledgments.} We would like to thank Guy Bencteux (EDF)
for valuable 
discussions and for his help in the implementation. C.L.B. and
E.C. would like to acknowledge many stimulating discussions with Richard
Lehoucq (Sandia National Laboratories).

%%%%%%%%%%%%%%%%%%%%%%%%%%%%%%%%%%%%%%%%%%%%%%%%%%%%%%%%%%%%%%%%%%%%%%%%%%

\newpage
 
\medskip

\medskip

\medskip

\medskip

\medskip

\medskip

\medskip

\medskip

\medskip

\medskip

\medskip

\medskip

\medskip

\medskip

\medskip

\end{document}